\documentclass[12pt]{article}

\usepackage{graphicx,amssymb,amsthm,amsmath,amsfonts,bm,color}
\usepackage[longnamesfirst]{natbib}
    
\usepackage{setspace}
\theoremstyle{plain}   \newtheorem{lemma}{Lemma} 

\theoremstyle{definition}   

\theoremstyle{remark}

\usepackage{hyperref}
\usepackage{bm}
\usepackage[ruled,vlined]{algorithm2e}
\usepackage{tabularx}       
\usepackage[table]{xcolor}
\definecolor{shade}{HTML}{F8F4FF} 

\newcommand{\STAR}{\textsc{star} }
\newcommand{\STARp}{\textsc{star}}

\usepackage[top=1in, right = 1in, left = 1in, lines = 25]{geometry}

\begin{document}

\title{\vspace{-15mm}  Semiparametric count data regression for \\ \vspace{-3mm} self-reported mental health \vspace{-4mm}}


\author{Daniel R. Kowal and Bohan Wu\thanks{
Department of Statistics, Rice University, Houston, TX (Correspondence: \href{mailto:Daniel.Kowal@rice.edu}{daniel.kowal@rice.edu}).}}
\date{}

\maketitle

\vspace{-10mm}
\begin{abstract} 
{\small ``For how many days during the past 30 days was your mental health not good?" The responses to this question measure self-reported mental health and can be linked to important covariates in the National Health and Nutrition Examination Survey (NHANES). However, these count variables present major distributional challenges: the data are overdispersed, zero-inflated, bounded by 30, and heaped in five- and seven-day increments. To address these challenges---which are especially common for health questionnaire data---we design a semiparametric estimation and inference framework for count data regression. The data-generating process is defined by \emph{simultaneously transforming and rounding} (\STARp) a latent Gaussian regression model. The transformation is estimated nonparametrically and the rounding operator ensures the correct support for the discrete and bounded data. Maximum likelihood estimators are computed using an EM algorithm that is compatible with \emph{any} continuous data model estimable by least squares. \STAR regression includes asymptotic hypothesis testing and confidence intervals, variable selection via information criteria, and customized diagnostics. Simulation studies validate the utility of this framework. Using \STAR regression, we identify key factors associated with self-reported mental health and demonstrate substantial improvements in goodness-of-fit compared to existing count data regression models.}
\end{abstract}

\thispagestyle{empty}

\clearpage
\setcounter{page}{1}

\section{Introduction}\label{intro}
The National Health and Nutrition Examination Survey (NHANES) asks a critical question:  ``For how many days during the past 30 days was your mental health not good?" The responses (\texttt{DaysMentHlthNotGood}) provide insights on self-reporting of mental health issues and can be linked to demographic, socioeconomic, behavioral, and health-related covariates.
Mental health and mental disorders are key factors in quality of life, depression, and risk of self-harm, and are focal points of many research studies   \citep{Scheid2017}. Previous research has sought to associate mental health, mental disorders, or depression with 
gender \citep{Seedat2009}, 
race \citep{Williams2010race}, 
socioeconomic status \citep{Ortega1990}, 
marital status \citep{Williams2010}, 
age \citep{Mirowsky1999}, 
smoking \citep{Klungsoyr2006}, 
and
blood pressure \citep{Tzourio1999}, among many other factors. Notably, NHANES data include relevant covariates for \emph{all} of these factors---and several others (see Table~\ref{tab:data})---on a large sample of individuals, and thus offers a unique opportunity for a joint analysis of multiple factors.

\begin{figure}[h]
\begin{center}
\includegraphics[width=.49\textwidth]{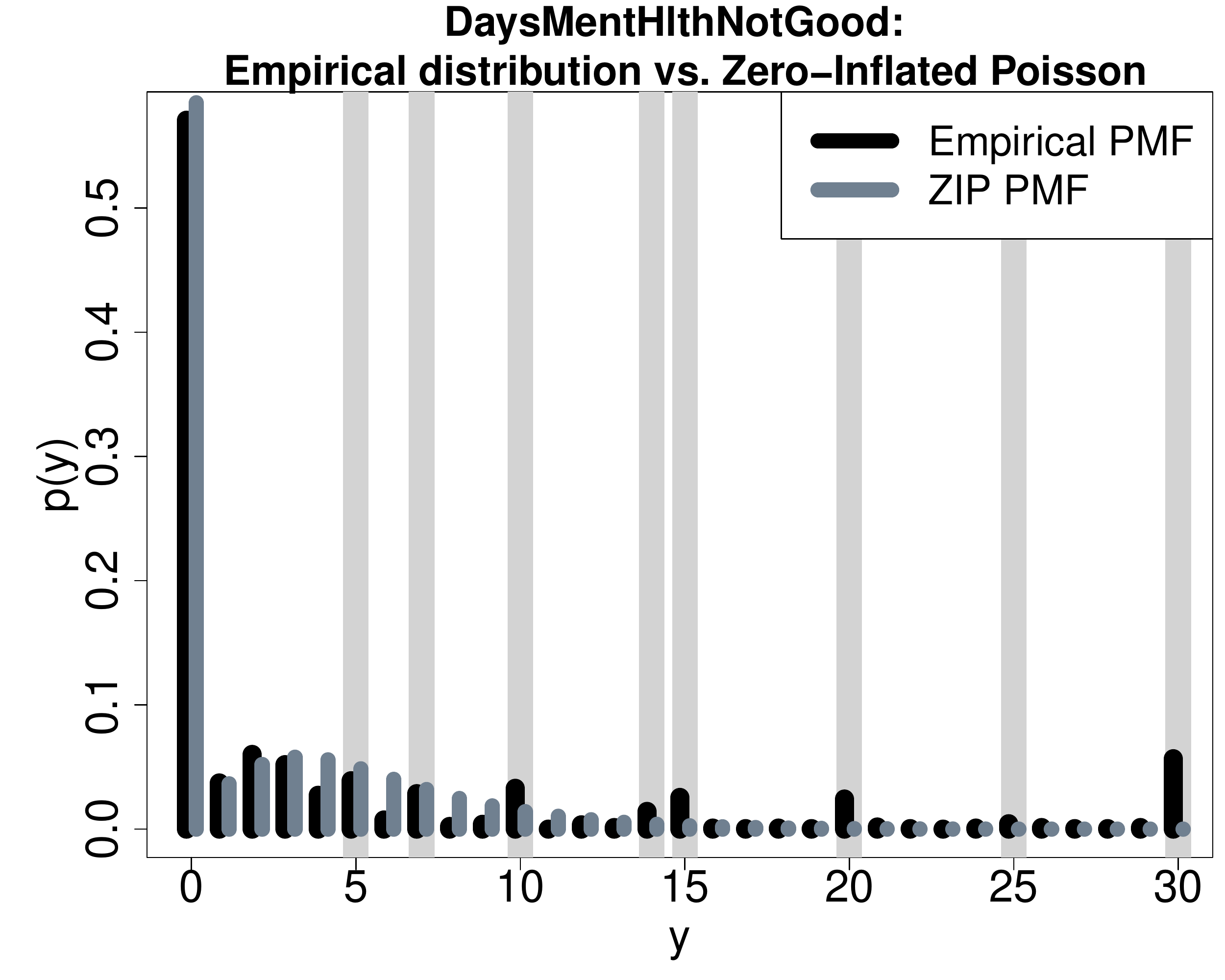}
\includegraphics[width=.49\textwidth]{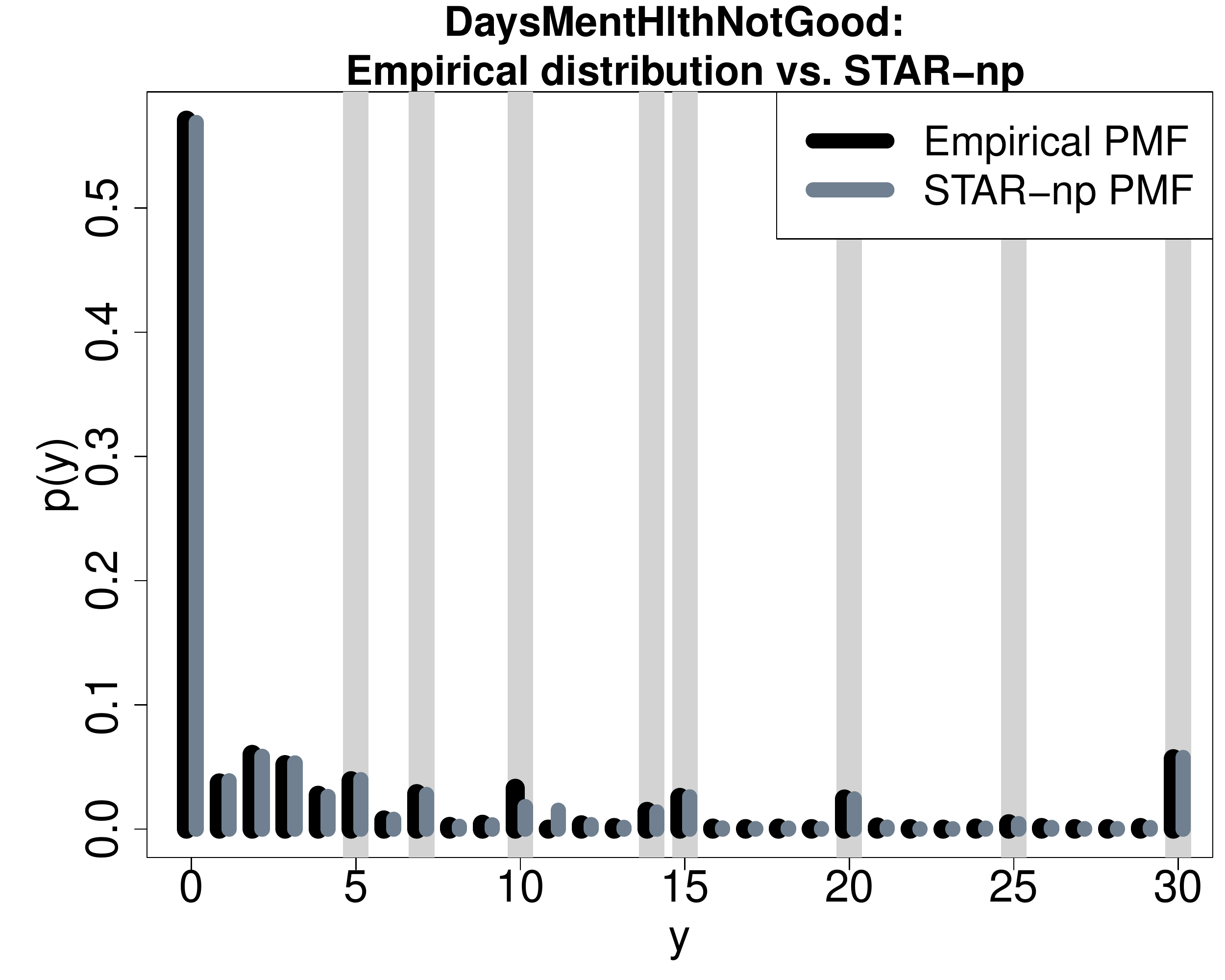}
\end{center}
\caption{\small Empirical probability mass function (PMF) for \texttt{DaysMentHlthNotGood} with estimated PMFs for zero-inflated Poisson (ZIP, left) and \STAR (right). 
While \STAR neatly captures zero-inflation, heaping (light gray), and boundedness, ZIP is inadequate for the nonzero counts. 
\label{fig:pred}}
\end{figure}

Our goal is to construct an adequate count regression model for these health questionnaire data and characterize the effects of covariates on self-reported mental health. Health questionnaire data often require customized statistical methodology, including life event stressors  \citep{Herring2004}, sleep quality \citep{Dunson2005}, nutritional and food intake \citep{Kipnis2009},  and drug use \citep{Song2017}, among many others. From a statistical modeling perspective, \texttt{DaysMentHlthNotGood} is an overdispersed, zero-inflated, and bounded count variable. Figure~\ref{fig:pred} shows the empirical probability mass function (PMF) for \texttt{DaysMentHlthNotGood}. The PMF has spikes at both the lower bound (zero days) and the upper bound (30 days) and, most uniquely, heaping in five- and seven-day increments. This is an expected consequence of \emph{self-reported} behavior: individuals are more likely to report 14 or 15 days than they are 16 or 17 days regardless of the exact truth. Hence, a regression model for \texttt{DaysMentHlthNotGood} must be capable of modeling  discreteness, overdispersion, zero-inflation, boundedness, and heaping.

Most count regression models build upon the Poisson distribution. Poisson regression can be suitable for point estimation or point prediction (see Section~\ref{sims}), but is often inadequate for modeling and inference due to the rigid equidispersion requirements. 
Extensions for over/underdispersion typically introduce additional parameters or latent variables, such as quasi-Poisson, Negative Binomial, or Conway-Maxwell-Poisson  \citep{sellers2010flexible}. Other features such as zero-inflation can be appended to the Poisson model or its generalizations. 
However, the added complexity of generalized Poisson models introduces computational challenges,  
undermines the interpretability of the regression model, and  often fails to produce an adequate model, especially for questionnaire data. Figure~\ref{fig:pred} illustrates this point for zero-inflated Poisson regression: the model is incapable of capturing the key features of \texttt{DaysMentHlthNotGood} and fails crucial residual diagnostic checks (see Figure~\ref{fig:qqplots}); similar results are observed for Negative Binomial models. 

An alternative strategy eschews count data models in favor of more familiar and flexible continuous data models. Indeed, some authors recommend applying a transformation to the count data (such as logarithmic or square-root) and proceeding with Gaussian linear models \citep{ives2015testing}; others recommend against this practice \citep{o2010not, st2018count} or provide strict guidelines under which it may be reasonable \citep{warton2016three}. This debate highlights two important yet conflicting points: continuous data models are often preferable---and occasionally  satisfactory---but are  fundamentally incoherent for count data and clearly cannot be expected to capture the key properties of \texttt{DaysMentHlthNotGood}. 

Informed by these benefits and limitations, we develop a semiparametric count regression model for health questionnaire data. The count data regression model is constructed by \emph{simultaneously transforming and rounding} (\STARp) a latent continuous data regression model. The latent regression model infuses the benefits of continuous data modeling while the rounding operator ensures a coherent integer-valued process. The transformation provides essential distributional flexibility and can be estimated nonparametrically via the marginal empirical cumulative distribution function (CDF). In conjunction, the transformation and rounding operations effectively map a count variable with challenging distributional features to a more convenient latent space for continuous regression modeling. This strategy allows the modeler to deploy interpretable and familiar continuous data regression models, such as Gaussian linear regression, for settings with difficult count-valued distributions. The effectiveness of the proposed approach for  \texttt{DaysMentHlthNotGood} is advertised in Figure~\ref{fig:pred}: the \STAR model captures the key features of \texttt{DaysMentHlthNotGood} and offers significant improvements relative to existing count regression models (see Section~\ref{apps}). 
Using likelihood-based inference, we provide a complete toolbox for count data regression: maximum likelihood estimators (MLEs) for estimation and prediction, hypothesis testing, confidence intervals, variable selection via information criteria, and customized diagnostics for \STAR regression.

The general framework of \STAR was initially developed by \cite{Kowal2020a} for Bayesian regression. Similarly, \cite{canale2011bayesian} and \cite{canale2013nonparametric} used rounded latent data models for nonparametric Bayesian inference and count data regression and \cite{Kowal2020target} applied \STAR for Bayesian functional data analysis. These methods demonstrate the promise of this modeling strategy, but  are exclusively Bayesian and do not consider frequentist estimation, inference, or diagnostics. In addition, these previous approaches primarily relied on fixed and known transformations, which cannot account for the heaping behavior that is common in questionnaire data. The proposed nonparametric model for the transformation resolves this issue and offers  superior results for the mental health questionnaire data (see Section~\ref{apps}).  Further, the proposed EM algorithm is compatible with any (weighted) least squares estimator, which immediately adapts many existing (non-Bayesian) machine learning models for continuous data to be more suitable for count-valued questionnaire data.



In Section~\ref{star}, we introduce the \STAR regression model; Section~\ref{estimation} develops the estimation and inference for \STAR regression; Section~\ref{apps} applies the \STAR regression model to the NHANES data; Section~\ref{sims} evaluates the proposed methods on simulated data; Section~\ref{discuss} concludes. 

\section{STAR models for count data}\label{star}
Consider the paired observations $\{\bm x_i, y_i\}_{i=1}^n$ of predictors $\bm x_i \in \mathcal{X} \subset \mathbb{R}^p$ and count response variables $y_i = y_i(\bm x_i) \in \mathcal{N}= \{0,1,\ldots,\infty\}$. Let $\mu_\theta \!:\mathcal{X}\to \mathbb{R}$ denote a real-valued regression function with parameters $\theta \in \Theta$, such as a linear model $\mu_\theta(\bm x) = \bm x'\bm\theta$.  \STAR seeks to link the regression function $\mu_\theta$ to the discrete data $y_i$ in a way that is interpretable, computationally convenient for estimation and inference, and well-defined for discrete data.  \STAR achieves these goals by introducing continuous latent random variables that are carefully connected to each other and the observable data via \emph{transformation} and \emph{rounding} operations. 

\STAR begins with a regression model for latent Gaussian random variables $z^*\!:\mathcal{X}\to \mathbb{R}$:
\begin{equation}\label{spn}
z_i^* = \mu_\theta(\bm x_i) +  \epsilon_i, \quad \epsilon_i \stackrel{iid}{\sim}N(0,\sigma^2), \quad i=1,\ldots,n.
\end{equation}
The latent data model \eqref{spn} is ideal: the regression function $\mu_\theta(\bm x_i) = \mathbb{E}(z_i^* | \bm x_i)$ is the conditional expectation of $z^*$ at $\bm x_i$ and the errors $\epsilon_i$ are iid Gaussian. If the latent data $z_i^*$ were observable, then the model parameters $\theta$ in \eqref{spn} could be estimated by maximum likelihood.

Instead, the latent data $z_i^*$ are linked to the observable data $y_i$ via transformation and rounding. The transformation step introduces a redundant latent variable $y_i^*\!: \mathcal{X}\to \mathcal{T}$ via a monotone transformation $g\!: \mathcal{T}\to \mathbb{R}$:
\begin{equation}\label{transform}
z_i^* = g(y_i^*).
\end{equation}
The transformation $g$  may be  known, such as logarithmic or square-root, or unknown and learned from the data (see Section~\ref{learn-g}). The purpose of $g$ is to provide modeling flexibility: the latent $y_i^*$ are transformed to $z_i^*$ and modeled as Gaussian. An appropriate choice of $g$ can improve the suitability of model \eqref{spn} for any regression function $\mu_\theta$. 

Lastly, the latent data $y_i^*$ are linked to the observed data via a rounding operation $h\!: \mathcal{T}\to \mathbb{N}$: 
\begin{equation}
\label{round}
y_i = h(y_i^*)
\end{equation}
where $h(y^*) = j \in \mathcal{N}$ when $y^* \in \mathcal{A}_j$ and $\{\mathcal{A}_j\}_{j=0}^\infty$ is a known partition of $\mathcal{T}$. We use rounding operators of the form $\mathcal{A}_j = [a_j, a_{j+1})$ for known $a_j \in \mathcal{T}$, which maps intervals to integers. The latent data $y_i^*$  provide a continuous proxy for the count data $y_i$, which are matched to the regression model \eqref{spn} through a transformation in \eqref{transform}. 
The coefficients $\theta$ in \eqref{spn} may be interpreted as in probit regression, which is a special case of \STARp: $\theta$ helps determine the likelihood for $y$, where larger values of $\mu_\theta(\bm x_i) $ imply greater probability on larger values of $y_i$.

The latent data $y_i^*$  are redundant: we may compress \eqref{transform}--\eqref{round} into
$y_i = h\{ g^{-1}(z_i^*)\}$
which, combined with \eqref{spn}, constitutes the \STAR regression model. From this perspective, the composition $h \circ g^{-1}$ is sufficient for mapping the Gaussian model \eqref{spn} to $\mathcal{N}$. We prefer to decouple $h$ and $g$ for several reasons. First, there exist many approaches for modeling a smooth and continuous function, which we can leverage to estimate the transformation $g$ (see Section~\ref{learn-g}). Second, $h$ can be specified to provide key distributional features, such as zero-inflation, boundedness, or censored data (see Lemma~\ref{star-prop}) without requiring any modifications to the estimation algorithm (see Section~\ref{em-alg}). Finally, the decoupled representation \eqref{spn}--\eqref{round} in insightful: \eqref{spn}--\eqref{transform} is comparable to the common practice of transforming the observed counts $y_i$ and modeling the transformed data as Gaussian, which is obtained by replacing $y_i^*$ with $y_i$ in \eqref{transform}. This approach implicitly acknowledges the utility of both the transformation and the continuous regression model. However, such models are not coherent for integer-valued data. The rounding operation in \eqref{round} corrects for this incoherence while preserving the core attributes of the transformed regression model.

The \STAR model not only incorporates a broad class of regression models in \eqref{spn}, but also offers vital distributional features for modeling count data:
 \begin{lemma}[\citealp{Kowal2020a}]\label{star-prop}
The \STAR model \eqref{spn}--\eqref{round} has the following properties:
\begin{enumerate}
\item \emph{Over/underdispersion:} for fixed $g$, there exist values of $(\mu_\theta, \sigma)$ such that $\mathbb{E}(y) < \mbox{Var}(y)$ (overdispersion) and  values of $(\mu_\theta, \sigma)$ such that $\mathbb{E}(y) > \mbox{Var}(y)$ (underdispersion).

\item \emph{Zero-inflation:} the probability of a zero count is 
$\mathbb{P}\{y(\bm x) = 0\} = \mathbb{P}\{z^*(\bm x) < g(a_1)\} = \Phi [ \{g(a_1) - \mu_\theta(\bm x_i)\}/\sigma]$.

\item \emph{Boundedness:} suppose there exists an upper bound $y_{max} \in \{1, \ldots, \infty\}$ for the counts $y$. By specifying the rounding operator such that $a_{y_{max} + 1} = \infty$ and requiring $\lim_{t \rightarrow\infty}g(t) = \infty$, we have  $\mathbb{P}(y \in \{0,1,\ldots, y_{max}\}) = 1$.

\item \emph{(Right) Censoring:} suppose only $y_i^c = \min\{y_i, C\}$ is observed and let  $\delta_i = \mathbb{I}\{y_i \ge C\}$ be a known censoring indicator. The likelihood for $\{y_i\}_{i=1}^n$ is 
$$L_c(\bm y; \theta, \sigma) = \prod_{i: \delta_i = 0} \mathbb{P}(y_i) \prod_{i: \delta_i = 1} \mathbb{P}(y_i \ge C) 
=  \prod_{i: \delta_i = 0} \mathbb{P}\{z_i^*\in [g(a_{y_i}), g(a_{y_i + 1}))\} \prod_{i: \delta_i = 1} \mathbb{P}\{z_i^* \ge g(a_C)\}
$$
which is \emph{identical} to the likelihood for $\{y_i^c\}_{i=1}^n$ subject to $y_{max} = C$ with $a_{y_{max} + 1} = \infty$ and $\lim_{t \rightarrow\infty}g(t) = \infty$. 
\end{enumerate}
\end{lemma}

The distributional features in Lemma~\ref{star-prop} are nontrivial. First, the capability to model over- or underdispersion immediately offers an advantage relative to Poisson models, which require $\mathbb{E}(y) = \mbox{Var}(y)$. Common generalizations of  Poisson models, such as Negative Binomial models, can capture overdispersion but not underdispersion. 
Next, zero-inflation simply corresponds to an abundance of small values of $z^*$. For example, if $g(t) = \log(t)$ and $a_j = j$, then $\mathbb{P}\{y(\bm x) = 0\}  = \mathbb{P}\{z^*(\bm x) < 0\}$, which is the probability of negative latent data in the regression model \eqref{spn}. \STAR is also compatible with a mixture-based approach for modeling zeros analogous to zero-inflated models; however, we have found that the baseline \STAR model \eqref{spn} is adequate for modeling an abundance of zeros in both real and simulated datasets. 

Among regression models for count data, the boundedness and censoring properties in Lemma~\ref{star-prop} are unique to \STARp. Boundedness and censoring are common in practice: for example, boundedness occurs for any count of daily events within a month ($y_{max} = 30$) and censoring occurs whenever data are recorded as ``$C$ or more" counts ($y_{max} = C$). Each property is easily incorporated into the model by setting $a_{y_{max} + 1} = \infty$ and requiring $\lim_{t \rightarrow\infty}g(t) = \infty$, which provides valid likelihood-based estimation and inference for bounded or censored data. No additional modifications to the estimation algorithm are required and the regression  specification in \eqref{spn} is not affected. By comparison, boundedness or censoring of Poisson and related models require a reformulation of the likelihood and customized estimation algorithms.

The latent representation for the \STAR model in \eqref{spn}--\eqref{round} produces a convenient (log-) likelihood with important properties for estimation. The PMF implied by \eqref{spn}--\eqref{round} is 
$
\mathbb{P}(y_i = j)= \mathbb{P}(y_i^* \in \mathcal{A}_j) = \mathbb{P}\{z_i^* \in g(\mathcal{A}_j)\} = \int_{g(\mathcal{A}_j)} \phi(z; \mu_\theta(\bm x_i), \sigma^2) \, dz
$
for $j \in \mathcal{N}$, where $\phi(\cdot)$ is the Gaussian density function implied by \eqref{spn}. For rounding operators of the form $\mathcal{A}_j = [a_j, a_{j+1})$, it follows that $g(\mathcal{A}_j) = [g(a_j), g(a_{j+1}))$ due to the monotonicity of $g$. Letting $\bm y = (y_1,\ldots, y_n)'$ denote the observed integer-valued data and noting the independence assumptions in \eqref{spn}, it follows that the log-likelihood is 
\begin{equation}\label{log-like}
\ell(\bm y; \theta, \sigma) 
= 
\sum_{i=1}^n \log \left[
\Phi\left\{\frac{g(a_{y_i + 1}) - \mu_\theta(\bm x_i)}{\sigma}\right\} - 
\Phi\left\{\frac{g(a_{y_i }) - \mu_\theta(\bm x_i)}{\sigma}\right\} 
 \right]
\end{equation}
where $\Phi$ denotes the CDF of a standard normal random variable. The log-likelihood does not include latent data: the \STAR model can be constructed directly from \eqref{log-like} without reference to $y_i^*$  or $z_i^*$. Evaluation of \eqref{log-like} is critical for subsequent hypothesis testing, confidence intervals, variable selection (via information criteria), and residual diagnostics (see Sections~\ref{estimation}~and~\ref{apps}).

Despite the apparent complexity of the log-likelihood in \eqref{log-like}, we obtain useful model properties and provide an efficient maximum likelihood estimation algorithm (Section~\ref{estimation}). For the \STAR linear regression model $\mu_\theta(\bm x) = \bm x ' \bm \theta$, we have the following important result:
\begin{lemma}\label{global-concave}
Suppose $\mu_\theta(\bm x_i) = \bm x_i'\bm\theta$ and $\bm X = (\bm x_1, \ldots, \bm x_n)'$ is full rank. The log-likelihood \eqref{log-like} is strictly concave in $(\bm \theta/\sigma, 1/\sigma)$.
\end{lemma}
The result follows immediately from \cite{pratt1981concavity}. The primary implication of Lemma~\ref{global-concave} is that MLEs for the \STAR linear model can be computed easily and efficiently using a variety of algorithms. In Section~\ref{estimation}, we introduce an EM algorithm to compute MLEs for \STAR models.  Lemma~\ref{global-concave} implies that the EM algorithm converges to the global MLE for \STAR linear models---and therefore avoids the local optima that often plague EM algorithms. Global concavity of the log-likelihood has other important implications for the MLEs:
\begin{lemma}\label{unique}
Under the conditions of Lemma~\ref{global-concave}, the MLEs for $(\bm \theta, \sigma)$ exist and are unique. 
\end{lemma}
Lemmas~\ref{global-concave}~and~\ref{unique} do not consider joint estimation of the transformation $g$ with the parameters $(\bm \theta, \sigma)$. These results hold under one of two conditions on $g$: either the transformation is known or the transformation is estimated prior to computing the MLEs (see Section~\ref{learn-g}). 

The log-likelihood in \eqref{log-like} resembles that of ordinal regression, which has been used for modeling count data \citep{valle2019ordinal}. However, there are important limitations of ordinal regression that are not present for \STARp. First, ordinal regression does not incorporate the numeric values of the data $y_i$ and instead only uses the ranks. That approach is reasonable for modeling ordered categories absent any numeric meaning, but ignores useful information available for count data. Second, ordinal regression introduces a latent cut point for every unique data value, each of which must be estimated. The analogous quantity in \STAR is the transformation $g$, which may be known or unknown, but readily incorporates smoothness to provide regularized estimation. Lastly, ordinal regression is only supported on the \emph{observed} values (or levels) of $y_i$ rather than $\mathcal{N}$, which is unnecessarily restrictive and suboptimal for predictive modeling.

\section{Estimation and inference for STAR models}\label{estimation}

\subsection{Data-driven determination of the transformation}\label{learn-g}
The transformation $g$ is an essential component of the \STAR model \eqref{spn}--\eqref{round}. Although the rounding operation \eqref{round} ensures a \emph{valid} count-valued process, it does not guarantee an \emph{adequate model} for count data. The transformation $g$ assists in this endeavor and provides the vital link between the continuous regression model  \eqref{spn} and the count-valued coherence of \eqref{round}. 

We propose to learn the transformation $g$ by linking it to the marginal CDF of $y$:
\begin{equation}\label{cdf}
F_y(j) = \mathbb{P}(y \le j) = \mathbb{P}\{z < g(a_{j+1})\} = F_z\{g(a_{j+1})\}
\end{equation}
where $F_z$ is the CDF of $z^*$ and $\mathcal{A}_j = [a_j, a_{j+1})$ for $j  \in \mathcal{N}$. Gaussianity is assumed for the latent $z^*$ in \eqref{spn} and determines $F_z$. Let  $z^* \sim N(\mu_z, \sigma_z^2)$ denote a working marginal distribution. The transformation $g$ is identifiable only up to location ($\mu_z$) and scale ($\sigma_z$). This parametrization will allow us to specify a location and scale of the transformation to match the first two moments of $z^*$ with those of $y$, which lends interpretability to the parameters $\mu_\theta$ and $\sigma$ in \eqref{spn}. Since $F_z(t) = \Phi\{(t - \mu_z)/\sigma_z\}$ for $t \in \mathbb{R}$, it follows from \eqref{cdf} that 
$g(a_{j+1}) = \mu_z + \sigma_z \Phi^{-1}\{F_y(j)\}.$ 
For the default case of $a_0 = -\infty$, $a_j = j$ for $j=1,\ldots,y_{max}$, and $a_{y_{max}+1} = \infty$, we define the transformation more broadly:
\begin{equation}\label{g-R}
g(t) = \mu_z + \sigma_z \Phi^{-1}\{F_y(t - 1)\}, \quad  t \in [1,   y_{max} + 1)
\end{equation}
with $g(t) = -\infty$ for $t<1$ and $g(t) = \infty$ for $ t \ge y_{max} + 1$. 
Similarly, we define the inverse to be $g^{-1}(s) = F_y^{-1}[\Phi\{(s - \mu_z)/\sigma_z\}] + 1$ and set $\lim_{s \rightarrow -\infty} g(s) = 0$ and $\lim_{s \rightarrow \infty} g(s) = y_{max} + 1$.

The representation in \eqref{g-R} suggests a simple estimation procedure. First, we match the marginal moments of $z^*$ and $y$: $\mu_z = \overline y$ is the sample mean and $\sigma_z = \widehat s_y$ is the sample standard deviation of $\{y_i\}_{i=1}^n$.   Next, we substitute a point estimate for the CDF of $y$ using the nonparametric estimator \begin{equation}\label{g-np0}
\widehat g_0(t) = \overline y + \widehat s_y \Phi^{-1}\{\widetilde F_y(t-1)\}, \quad t \in [1, y_{max} + 1)
\end{equation}
where $\widetilde F_j(j) = \frac{n}{n+1} \widehat F_y(j)$ is a rescaling of the empirical CDF  $\widehat F_y(j) = \frac{1}{n}\sum_{i=1}^n \mathbb{I}\{y_i \le j\}$ to avoid boundary issues. As in \eqref{g-R}, we set $\widehat g_0(t) = -\infty$ for $t<1$ and $\widehat g_0(t) = \infty$ for $t \ge y_{max}+1$. Related strategies to estimate a transformation using the empirical CDF have been used for Gaussian graphical models  \citep{Liu2009} and linear regression \citep{Siegfried2020}. However, these methods do not directly consider zero-inflation, boundedeness/censoring, or heaping. Parametric alternatives, such as Poisson or Negative Binomial CDFs, offer the ability to conform \STAR with a pre-specified marginal distribution for the counts. In this case, the accompanying parameters of $F_y$ must be estimated. By default, we use method-of-moment estimators that recycle $\overline y$ and $\widehat s_y$. 

The transformation  \eqref{g-np0} resembles a classical Gaussian copula model. In fact, $\widehat g_0$ eliminates the need for the rounding operator $h$: since $F_y(t) = F_y(\lfloor t \rfloor )$ for $y \in \mathcal{N}$,  $\widehat g_0(t) = \widehat g_0(\lfloor t \rfloor )$ is a step function with integer-valued inverse $\widehat g_0^{-1}(z^*) \in \mathcal{N}$ for $z^* \in \mathbb{R}$. The critical drawback of \eqref{g-np0} is that the data-generating process defined by $\widehat g_0^{-1}(z^*)$ for Gaussian $z^*$ can \emph{only} generate the observed values $\{y_i\}_{i=1}^n$. Therefore, any value in $\mathcal{N}$ that is not observed in the data is also not supported by the \STAR model with transformation \eqref{g-np0}: $\mathbb{P}(y = j) = 0$ whenever $g(a_j) = g(a_{j+1})$, which occurs whenever $j \not \in \{y_i\}_{i=1}^n$. This restriction is unnecessary and can inhibit prediction.

To remedy this issue, we again decouple the rounding and transformation functions. We compute a smooth, monotonic interpolation of $\widehat g_0$ on the observed data values $\{y_i\}_{i=1}^n$: 
\begin{equation}\label{interpo}
\widehat g(y_i) = \widehat g_0(y_i), \quad i=1,\ldots,n
\end{equation}
where $\widehat g$ is a monotonic cubic spline estimated using the \texttt{R} function \texttt{splinefun(..., method = `monoH.FC')}  \citep{Fritsch1980},  which does not require any tuning parameters. The limits $\widehat g(t) = -\infty$ for $t<1$ and $\widehat g(t) = \infty$ for $t \ge y_{max}+1$ are also imposed. The estimate $\widehat g$ interpolates $\widehat g_0$ at each increment, but connects these points smoothly and monotonically to induce a data-generating process  supported wholly on $\{0,\ldots, y_{max}\}$. Similarly,  $\widehat g_0$ is bounded, since  $\widetilde F_y(t) = n/(n+1)$ for any $t \ge \max y_i$, while the cubic interpolating spline $\widehat g$ is linear (and  increasing) beyond this boundary.  As a result, $\widehat g$ preserves the best of both worlds: the smoothness appropriately expands the support of the data-generating process, while the interpolation \eqref{interpo} ensures that $\widehat g$ recovers the same CDF as $\widehat g_0$ at the observed data values $\{y_i\}_{i=1}^n$.  Lastly, since $\widehat g$ is estimated in advance of $\theta$ and $\sigma$, the log-likelihood concavity results from Lemma~\ref{global-concave} are unaffected. We refer to the \STAR model with this nonparametric estimator of $g$ as \STARp-np.

For Bayesian \STAR models, \cite{Kowal2020a} proposed Box-Cox transformations $g(t ; \lambda) = (t^\lambda - 1)/\lambda$ with $\lambda > 0$ and $g(t; \lambda =0) = \log(t)$.  Important special cases include the (shifted) identity transformation $\lambda = 1$, the (shifted and scaled) square-root transformation $\lambda = 1/2$, and the log-transformation. Box-Cox functions are popular for transforming continuous data towards Gaussianity, which in the present setting corresponds to \eqref{spn}--\eqref{transform}. However, these transformations impose a restrictive parametric form, often lack interpretability, and require selection or estimation of $\lambda$. Nonetheless, Box-Cox transformations are compatible with the proposed approach,   and, along with the aforementioned parametric (Poisson or Negative Binomial) CDFs, may be  preferable to \STARp-np  when $n$ is small.

\subsection{An EM algorithm for maximum likelihood estimation}\label{em-alg} 

We introduce an EM algorithm \citep{dempster1977maximum} for computing MLEs under the \STAR model. 
EM algorithms use iterative optimizations that exploit the computational convenience of a \emph{complete data} likelihood. For \STARp, we define the complete data to be $(\bm y, \bm z^*)$ with latent data $\bm z^* = (z_1^*,\ldots, z_n^*)'$. For simplicity, let $\theta$ denote all unknown parameters, including $\sigma^2$. The complete data log-likelihood for \STAR is
\begin{align}
\label{complete-1}
\log p(\bm y, \bm z^* | \theta) &=  \log p(\bm y | \bm z^*, \theta) + \log p(\bm z^* | \theta)\\
\label{complete-3}
&= C(\bm y, \bm z^*) - \frac{1}{2} \sum_{i=1}^n \log(\sigma^2) - \frac{1}{2} \sum_{i=1}^n \Big[\{z_i^* - \mu_\theta(\bm x_i)\}^2/\sigma^{2}\Big]
\end{align}
where $C(\bm y, \bm z^*)$ is a constant that does not depend on $\theta$.  
%
The EM algorithm iteratively updates the parameters $\theta$ by  alternating the following steps at each iteration $s$:
\begin{itemize}
\item[] {\bf E-Step:} Compute  $Q(\theta, \theta^s) = \mathbb{E}_{[z^* | y, \theta = \theta^s]}\log p(\bm y, \bm z^* | \theta)$, where the expectation is taken with respect to the distribution $[\bm z^* | \bm y, \theta]$ for $\theta = \theta^s$ the parameter value at iteration $s$;
\item[] {\bf M-Step:} Maximize the conditional expectation, $\theta^{s+1} = \arg\max_{\theta} Q(\theta, \theta^s)$.
\end{itemize}
The algorithm proceeds until convergence, which is typically assessed by monitoring the stability of the log-likelihood \eqref{log-like} across iterations. 
Note that although the EM algorithm always produces non-decreasing log-likelihoods across iterations $s=1,\ldots,S$, in general the algorithm is not guaranteed to converge to a global maximum. However, for  \STAR  linear models (Lemma~\ref{global-concave}) and other special cases, the EM algorithm will converge to the unique global MLEs.

Under the \STAR model, computation of $Q(\theta, \theta^s)$ is straightforward:
\begin{eqnarray} \label{estep}
Q(\theta, \theta^s) &= & C(\bm y) - \frac{1}{2} \sum_{i=1}^n \log(\sigma^2)- \frac{1}{2} \sum_{i=1}^n \mathbb{E}_{[z^* | y, \theta = \theta^s]}\left[\{z_i^* - \mu_\theta(\bm x_i)\}^2/\sigma^{2}\right] \nonumber \\
&=& C(\bm y) - \frac{n}{2} \log(\sigma^2)
- \frac{1}{2\sigma^2}\left\{ \sum_{i=1}^n  \widehat z_i^{(2)} 
-2  \sum_{i=1}^n \mu_\theta(\bm x_i)\widehat z_i^{(1)} 
+ \sum_{i=1}^n \mu_\theta(\bm x_i)^2 \right\}
\end{eqnarray}
 where $C(\bm y)$ is a constant that does not depend on $\theta$ and $\widehat z_i^{(k)}  =\mathbb{E}\left\{(z_i^*)^k | y, \theta = \theta^s\right\}$, $k=1,2$ is the $k$th moment of the distribution $\left[z_i^*| \bm y,  \theta = \theta^s\right]$. 
 Under model \eqref{spn}, this distribution is  $\left[z_i^*| \bm y, \theta\right] \stackrel{indep}{\sim} N(\mu_\theta(\bm x_i), \sigma^2)$ truncated to $g(\mathcal{A}_{y_i})$ and using the parameter values $\theta = \theta^s$. For the common special case of $\mathcal{A}_j = [a_j, a_{j+1})$,   the required moments are computable as follows:
 \begin{align}
 \label{m1}
 \widehat z_i^{(1)}  &= \mu_\theta(\bm x_i) + \sigma \frac{\phi(a_i) - \phi(b_i)}{\Phi(b_i) - \Phi(a_i)} \\
  \label{m2}
 \widehat z_i^{(2)} &=  \mu_\theta(\bm x_i)\left\{\mu_\theta(\bm x_i) + 2 \sigma \frac{\phi(a_i) - \phi(b_i)}{\Phi(b_i) - \Phi(a_i)}\right\} + \sigma^2 \left\{
 1 + \frac{a_i\phi(a_i) - b_i\phi(b_i)}{\Phi(b_i) - \Phi(a_i)}
 \right\}
 \end{align}
 where $\phi$ is the standard normal density function, $a_i = \{g(a_{y_i}) - \mu_\theta(\bm x_i)\}/\sigma$, and $b_i = \{g(a_{y_i + 1}) - \mu_\theta(\bm x_i)\}/\sigma$. Note that the parameters in \eqref{m1}--\eqref{m2} will update at each iteration $s$ in the M-step.  
 
The form of \eqref{estep} produces important simplifications for the M-step: 
\begin{lemma}\label{EM-gen}
Consider a \STAR model \eqref{spn} generalized to include heteroskedastic errors, $\epsilon_i \stackrel{indep}{\sim}N(0,\sigma^2/w_i)$, where $\theta \in \Theta$ and $\sigma^2 > 0$ are the unknown parameters and $w_i > 0$ are known weights. 
  The M-step solutions for $(\theta, \sigma^2)$ are $\widehat \theta = \widehat \theta( \{\bm x_i,  \widehat z_i^{(1)}\}_{i=1}^n)$ and \begin{equation}
\label{mle-sig}
\widehat \sigma^2 = \frac{1}{n}\Big\{\sum_{i=1}^n w_i \widehat z_i^{(2)}  + \sum_{i=1}^n  w_i \mu_{\widehat \theta}^2(\bm x_i) -2 \sum_{i=1}^n w_i \mu_{\widehat \theta}(\bm x_i)\widehat z_i^{(1)} \Big\},
\end{equation}
where $\widehat \theta( \cdot)$ is the weighted least squares solution to 
\begin{equation}\label{sse}
\widehat \theta( \{\bm x_i, d_i\}_{i=1}^n) = \arg\min_{\theta \in\Theta} \sum_{i=1}^n w_i \{d_i - \mu_\theta(\bm x_i)\}^2
\end{equation}
for paired observations $(\bm x_i, d_i)$ with $d_i \in \mathbb{R}$ and $w_i  > 0$.
\end{lemma}
The proof is in the supplementary material, along with a complete  description of the EM algorithm.  For \emph{any} model that can be estimated using weighted least squares, Lemma~\ref{EM-gen} implies that the same model can be adapted for integer-valued data using the \STAR model \eqref{spn} and that the algorithm solving the least squares problem \eqref{sse} is sufficient for 
estimation in \STARp.

To compute fitted values from \STARp, we note that the expected value of the response is 
\begin{equation}\label{exp-count}
\mathbb{E}\{y(\bm x)\} = \sum_{j=0}^{y_{max}} j \mathbb{P}\{y(\bm x) = j\} = \sum_{j=0}^{y_{max}} j \Big\{\Phi\Big( \frac{g(a_{j+1}) - \mu_\theta(\bm x)}{\sigma}\Big) -  \Phi\Big( \frac{g(a_{j}) - \mu_\theta(\bm x)}{\sigma}\Big)\Big\}
\end{equation}
and is a function of $\theta$ and $\sigma$. The invariance property of MLEs ensures an MLE for $\mathbb{E}\{y(\bm x)\}$:
\begin{lemma}\label{mle}
The maximum likelihood estimator of $\mathbb{E} \{y(\bm x)\}$ conditional on $\bm x \in \mathcal{X}$ is 
$\widehat y(\bm x) \equiv \sum_{j=0}^{y_{max}} j\Big[\Phi\Big\{ \frac{g(a_{j+1}) -  \mu_{\widehat \theta}(\bm x)}{\widehat\sigma}\Big\}-  \Phi\Big\{ \frac{g(a_{j}) - \mu_{\widehat \theta}(\bm x)}{\widehat\sigma}\Big\}\Big]
,$ 
where $\widehat \theta$ and $\widehat \sigma$ are MLEs for $\theta$ and $\sigma$, respectively.
\end{lemma}
Extensions for heteroskedastic errors as in Lemma~\ref{EM-gen} are straightforward. In practice, we truncate the summation at $J(\bm x)$ pointwise for each $\bm x$, where $J(\bm x)$ is the 99.99th quantile of the distribution of $y(\bm x)$ and computed as $h[g^{-1}\{z_q^*(\bm x)\}]$ where $z_q^*(\bm x)$ is the $q$th quantile of the distribution of $z^*(\bm x)$.  The computing time of $\widehat y(\bm x)$ increases with $J(\bm x)$, but so  does the accuracy.

The computational  scalability of the EM algorithm depends primarily on  the least squares solution \eqref{sse} and the number  of iterations to convergence. Least squares solutions are typically highly scalable, especially for linear models. For the  data analysis in Section~\ref{apps},  convergence (with a  log-likelihood tolerance of  $10^{-10}$) occurs in about 40 iterations.

\subsection{Inference for STAR models}\label{inference}
We develop likelihood-based inference for the \STAR model \eqref{spn}. Under classical regularity conditions, which are satisfied for the linear regression model $\mu_\theta(\bm x) = \bm x' \bm\theta$, the \STAR MLEs are consistent and admit generalized likelihood ratio tests for asymptotic inference. 
For the composite hypothesis test $H_0: \theta \in \Theta_0$ against the alternative $H_1: \theta \in \Theta \setminus \Theta_0$ for $\theta \in \Theta$ and $\Theta_0 \subset \Theta$, the likelihood ratio test statistic is 
$\Lambda(\bm y) = \sup_{\theta \in \Theta_0} p(\bm y | \theta)/\sup_{\theta \in \Theta} p(\bm y | \theta)$,
where $p(\bm y|\theta)$ denotes the likelihood. The test statistic $\Lambda(\bm y)$ requires the global MLE for $\theta \in \Theta$, the restricted MLE for $\theta \in \Theta_0$, and the likelihood evaluated at these MLEs. Under \eqref{spn}, the log-likelihood \eqref{log-like} is sufficient. The rejection region for the likelihood ratio test is defined as $\{\bm y: \Lambda(\bm y) \le c\}$ for some $c \in [0,1]$. 

The asymptotic distribution of the likelihood ratio test statistic is known: under the null hypothesis, $-2 \log \Lambda(\bm y) \sim \chi_r^2$ as $n \rightarrow \infty$, where $r = \dim \Theta - \dim \Theta_0$ \citep{wilks1938large}. The rejection region for an $\alpha$-level test is therefore $-2 \log \Lambda(\bm y) >  c_{\alpha, r}$, where $c_{\alpha, r}$ is the upper $\alpha$-quantile of the $\chi^2$-distribution with $r$ degrees of freedom. Given MLEs $\widehat \theta_0$ and $\widehat \theta$ computed over the sets $\Theta_0$ and $\Theta$, respectively, and for sufficiently large sample size $n$, we may therefore test composite hypotheses  under the \STAR model.

The likelihood ratio test statistic also provides a mechanism for constructing a confidence set for $\theta$. In particular, a $(1-\alpha)$ confidence set for $\theta$ is  
$C(\bm y) = \{ \theta_0: \bm y \in A(\theta_0)\}$,  where $A(\theta_0)$ is the acceptance region of an $\alpha$-level  test of $H_0: \theta = \theta_0$ (\citealp{casella2002statistical}, Theorem 9.2.2). For each $\theta_0$, the acceptance region for the (asymptotic) $\alpha$-level test is 
$\log p(\bm y | \widehat \theta) - \log p(\bm y | \theta_0) < c_{\alpha, 1}/2$
which implies that the (asymptotic) confidence set is equivalently
$C(\bm y) = \big\{ \theta_0: \log p(\bm y|\theta_0) >  \log p(\bm y | \widehat \theta) - c_{\alpha, \dim(\theta) - 1}/2\big\}.$
This confidence set  may be constructed without computing derivatives of the log-likelihood \eqref{log-like}, which, although straightforward in general, would require case-specific derivations depending on the relationship between the conditional mean $\mu_\theta$ and the unknown parameters $\theta$. Note that this approach for constructing confidence intervals is identical to that used for non-Gaussian GLMs in \texttt{R} via the \texttt{MASS} package.

 \subsection{Model diagnostics}
Model diagnostics typically rely on inspection of the residuals from the fitted model. For non-Gaussian models, various options for residuals are available, such as Pearson or deviance residuals \citep{mccullagh1989generalized}. However, for discrete data, standard residuals diagnostics are often misleading due to the nonuniqueness of the observations $\{y_i\}_{i=1}^n$. \cite{dunn1996randomized} proposed a randomization-based procedure that produces continuous and Gaussian residuals, which improves diagnostic assessment. Letting $U^{(0,1)} \sim \mbox{Uniform}(0,1)$, Dunn-Smyth residuals for \STAR are given by $r_i = \Phi^{-1}(u_i)$, where $ u_i = \mathbb{P}\{y(\bm x) = y_i\}U^{(0,1)} + \mathbb{P}\{y(\bm x) < y_i\} = \Big[\Phi \Big\{ \frac{g(a_{y_i+1}) - \mu_{\widehat \theta}(\bm x_i)}{\widehat \sigma}\Big\} -  \Phi\Big\{ \frac{g(a_{y_i}) - \mu_{\widehat \theta}(\bm x_i)}{\widehat \sigma}\Big\}\Big]U^{(0,1)} + \Phi\Big\{ \frac{g(a_{y_i - 1}) - \mu_{\widehat \theta}(\bm x_i)}{\widehat \sigma}\Big\}$
is readily computable based on the MLEs $\widehat \theta$ and $\widehat \sigma$. We recommend examining multiple instances of $r_i$ to verify that conclusions are not specific to a particular realization of $U^{(0,1)}$.

\section{NHANES data analysis}\label{apps}
Using data from the 2011--2012 National Health and Nutrition Examination Survey (NHANES), we study the demographic, socioeconomic, behavioral, and health-related factors that predict self-reported mental health (\texttt{DaysMentHlthNotGood}). 
The covariates ($p=23$) are summarized in Table~\ref{tab:data}. Individuals with missing,  refused, or ``don't know" responses are excluded; additional pre-processing details are provided in the supplementary material.  The analysis dataset includes $n=2311$ individuals.
We fit the \STAR linear model to these data using the smoothed empirical CDF from Section~\ref{learn-g} (\STARp-np) and the Box-Cox transformation with fixed $\lambda = 1/2$ (\STARp-sqrt). In addition, we include Gaussian regression on $\log(y+1)$ (Gauss-log), Poisson and zero-inflated Poisson (ZIP) regression,  Negative Binomial (NegBin) and zero-inflated Negative Binomial (ZINB) regression, and a logistic regression model that incorporates the upper bound 30 as the number of ``trials" in the binomial distribution (Binom).

Since NHANES data are collected by a complex multistage sampling design, we include several mechanisms to adjust for or quantify the effects of the sampling design. Following \cite{Lumley2017}, we include the primary sampling unit as a covariate to account for possible geographic clustering effects. In addition, the variables that determine the survey weights---gender, age, and race---are all included, which can provide a model-based adjustment to the design \citep{Gelman2007,Niu2019}. 
For comparison, we compute MLEs for the heteroskedastic \STARp-np model from Lemma~\ref{EM-gen} using the individual survey weights for $w_i$, which provides a design-consistent estimator for the \STAR regression coefficients via maximization of a \STAR pseudo-likelihood \citep{Lumley2017}. In this case, we also include the survey weights for the constituent quantities of the estimated transformation $\widehat g$: the sample mean, sample standard deviation, and empirical CDF of $y$.  Note that for general use of \STARp, consideration of the weighting is typically not necessary. 

To summarize the competing regression models, we compute the -2 log-likelihood for each model in Table~\ref{tab:loglike}. The \STAR models outperform all competitors by a wide margin.   Although the models differ in the number of parameters,  we note that \STARp-sqrt contains the same number of parameters as Gauss-log and NegBin yet fewer parameters than the zero-inflated models. Since $y \in \{0,1,\ldots, 30\}$ only admits 31 unique values, the smoothed empirical CDF for \STARp-np counts no more than 31 additional parameters compared to \STARp-sqrt. As a result, it is straightforward to verify that AIC and BIC rank \STARp-np first and \STARp-sqrt second ahead of the competing models.

\begin{table}[h]
\centering
\begin{tabular}{rrrrrrrr}
  \hline
Gauss-log & Poisson & ZIP & NegBin & ZINB & Binom & \STARp-np & \STARp-sqrt \\ 
  \hline
  11025 & 26208 & 14391 & 9636 & 9367 & 34321 & {\bf 8146} & 8691 \\ 
   \hline
\end{tabular}
\caption{\small -2 log-likelihood for each regression model. The \STAR models are clear favorites by likelihood metrics, including AIC and BIC. Note that Gauss-log includes a change-of-variables to the original data scale.\label{tab:loglike}}
\end{table}

\begin{figure}[h!]
\begin{center}
\includegraphics[width=.49\textwidth]{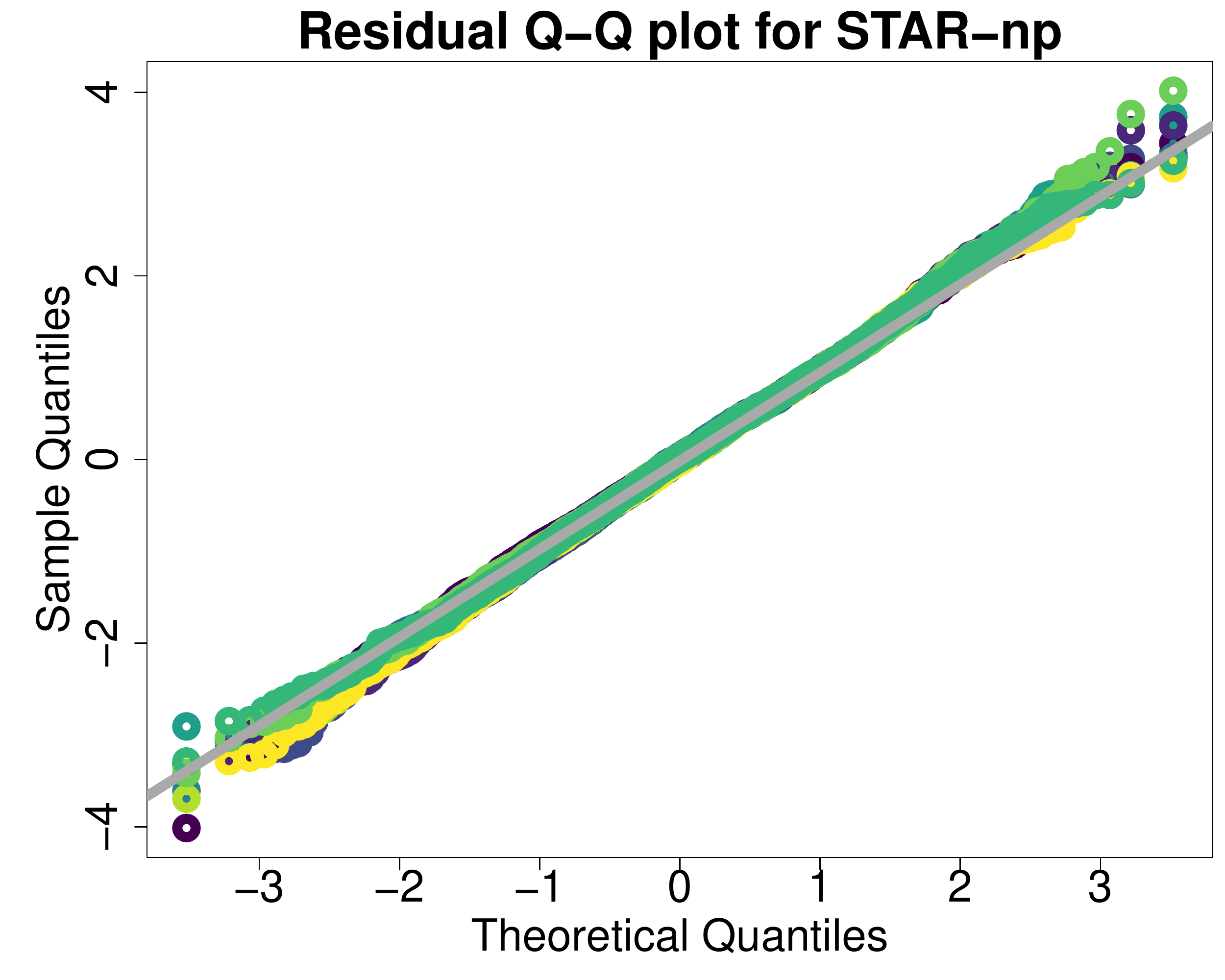}
\includegraphics[width=.49\textwidth]{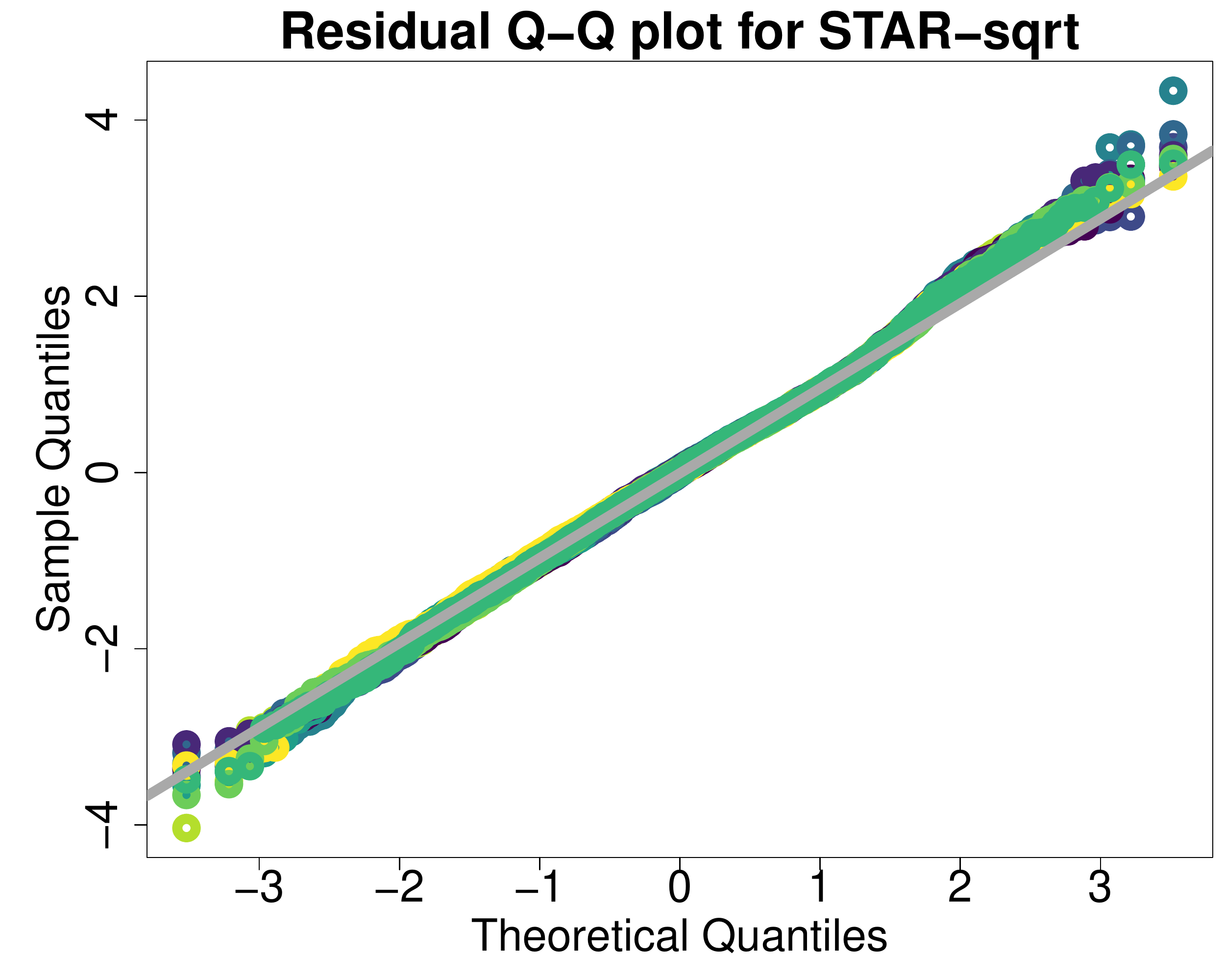}
\includegraphics[width=.49\textwidth]{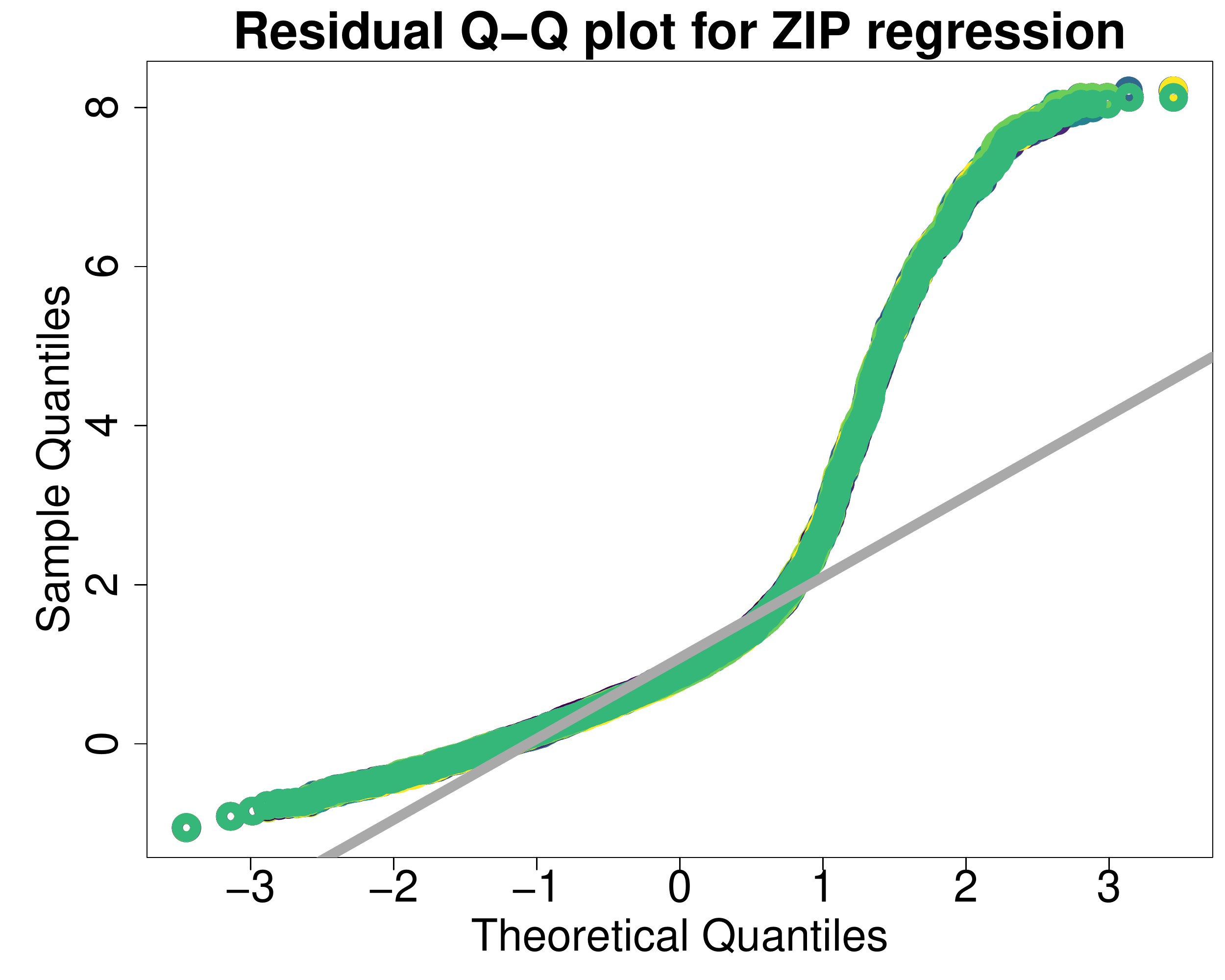}
\includegraphics[width=.49\textwidth]{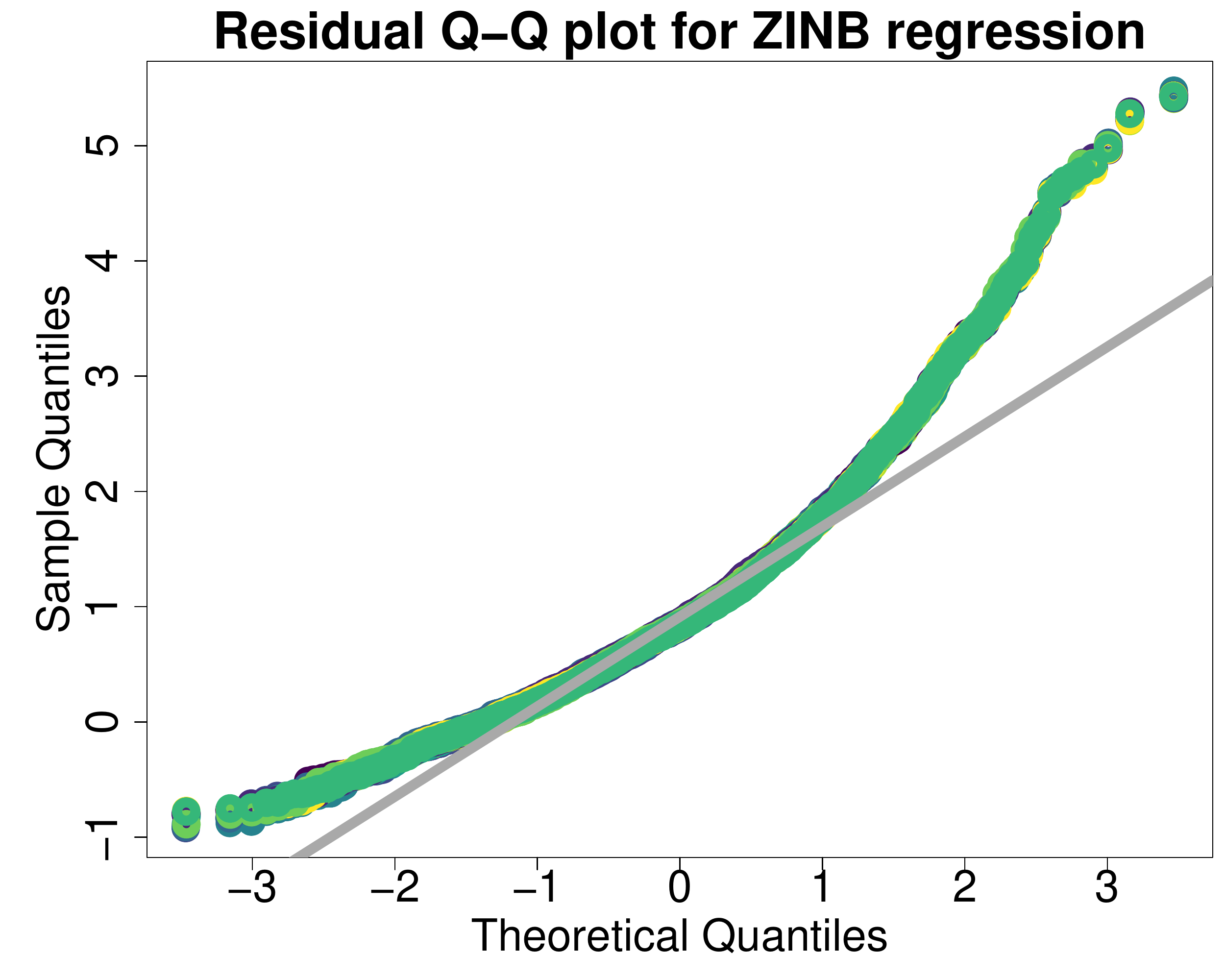}
\end{center}
\caption{\small  Dunn-Smyth residual QQ-plots for \STARp-np (top left), \STARp-sqrt (top right),  ZIP (bottom left), and ZINB (bottom right). Points along the line indicate goodness-of-fit. Each plot includes 10 sets of the randomized residuals (color-coded).  Both \STAR models provide excellent fits, while the ZIP and ZINB models are clearly inadequate.
\label{fig:qqplots}}
\end{figure}

Visually, the preference for \STAR is even more striking. Figure~\ref{fig:qqplots} presents the Dunn-Smyth residual diagnostics for \STARp-np, \STARp-sqrt, and ZIP. Most notably, \STARp-np provides an excellent fit to the data. More surprising, \STARp-sqrt---which cannot account for heaping---also provides a reasonable fit. Despite the simplicity of the transformation in \STARp-sqrt, the rounding operator crucially incorporates both zero-inflation and boundedness ($y_{max} = 30$). By comparison, the ZIP model accounts only for zero-inflation and is decidedly inadequate for these data.

\begin{figure}[h!]
\begin{center}
\includegraphics[width=1\textwidth]{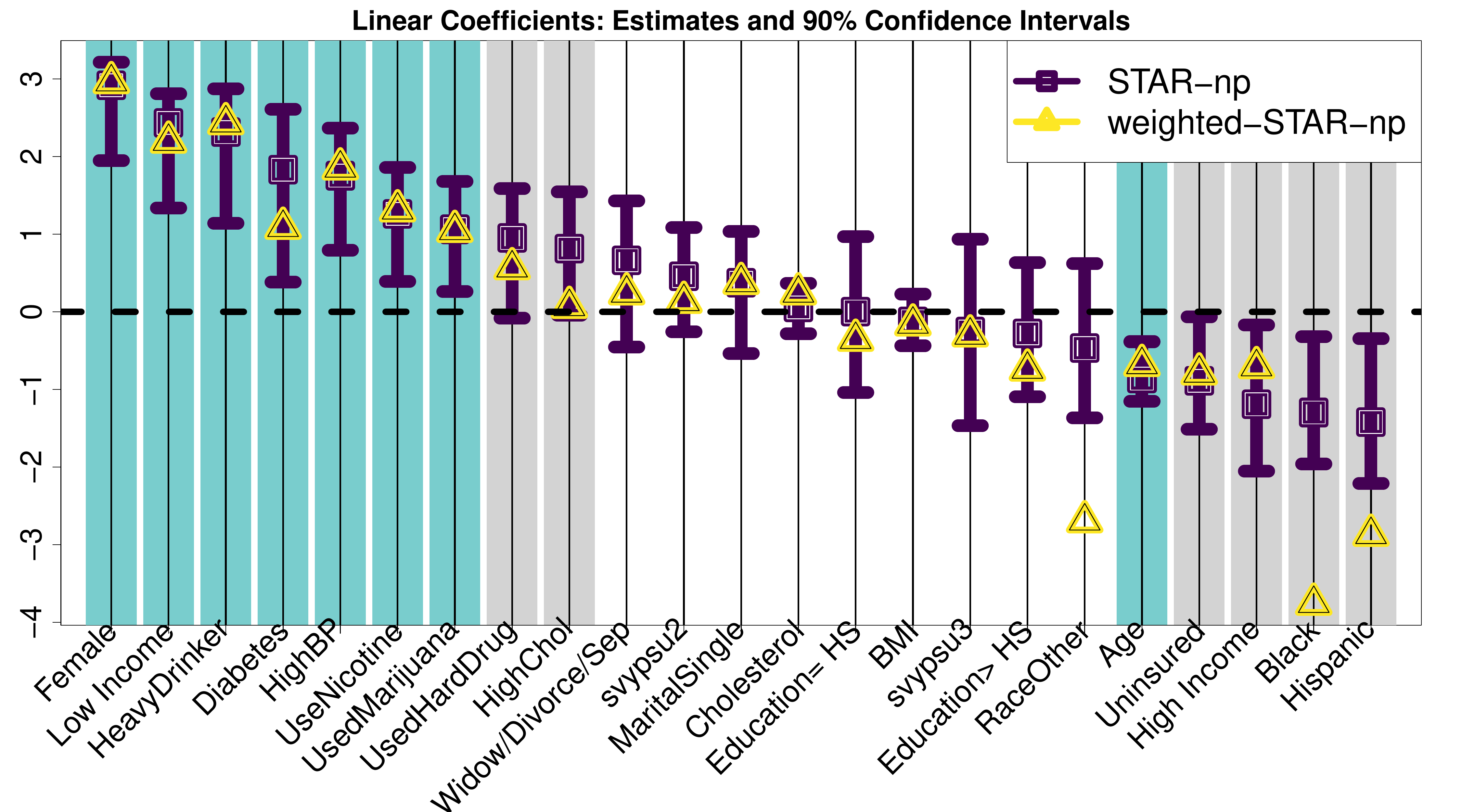}
\end{center}
\caption{\small Estimated coefficients and 90\% confidence intervals for \STARp-np (blue) along with the survey-weighted point estimates (yellow). The BIC-selected variables (blue) and AIC-selected variables (blue and gray) are annotated. The survey-adjusted estimates are similar to the unweighted estimates with the exception of the race categories, each of which is oversampled in the NHANES design. 
\label{fig:coef}}
\end{figure}

Proceeding with \STARp-np, we summarize the results of the fitted model in Figure~\ref{fig:coef}. The \STARp-np MLEs and 90\% confidence intervals for each coefficient present the magnitude and direction of each variable's effect with uncertainty quantification. Using a backward-elimination algorithm, we select (and annotate) variables according to AIC and BIC, both of which use the \STARp-np likelihood. The selected variables with (directionally) positive effects on self-reported poor mental health include demographic (female) and socioeconomic  (low income) variables, behavioral attributes (heavy drinking, recent nicotine use, any use of marijuana, any use of hard drugs), and health variables (diabetes and high blood pressure). The only selected variable with a (directionally) negative effect is age, but the estimated effects of uninsured, Black, Hispanic, and high income status are also negative. Moreover, the survey-weighted \STARp-np MLEs help assess the effect of the sampling design on the point estimates. The estimates are similar with the exception of the race categories, each of which is oversampled in the NHANES design.

To target the effect of drug use, we compute a \STAR likelihood ratio test  comparing the full \STARp-np model against the reduced \STARp-np model that excludes nicotine, marijuana, or hard drug use. The likelihood ratio test obtained a $p$-value $1.45\times 10^{-6}$, providing strong evidence that drug use is an important factor in self-reported mental health. The estimated effect of each variable is positive  (Figure~\ref{fig:coef}), which suggests that drug use is associated with more self-reported days of poor mental health---even after adjusting for other demographic, socioeconomic,  behavioral, and health-related factors.

\begin{figure}[h!]
\begin{center}
\includegraphics[width=.6\textwidth]{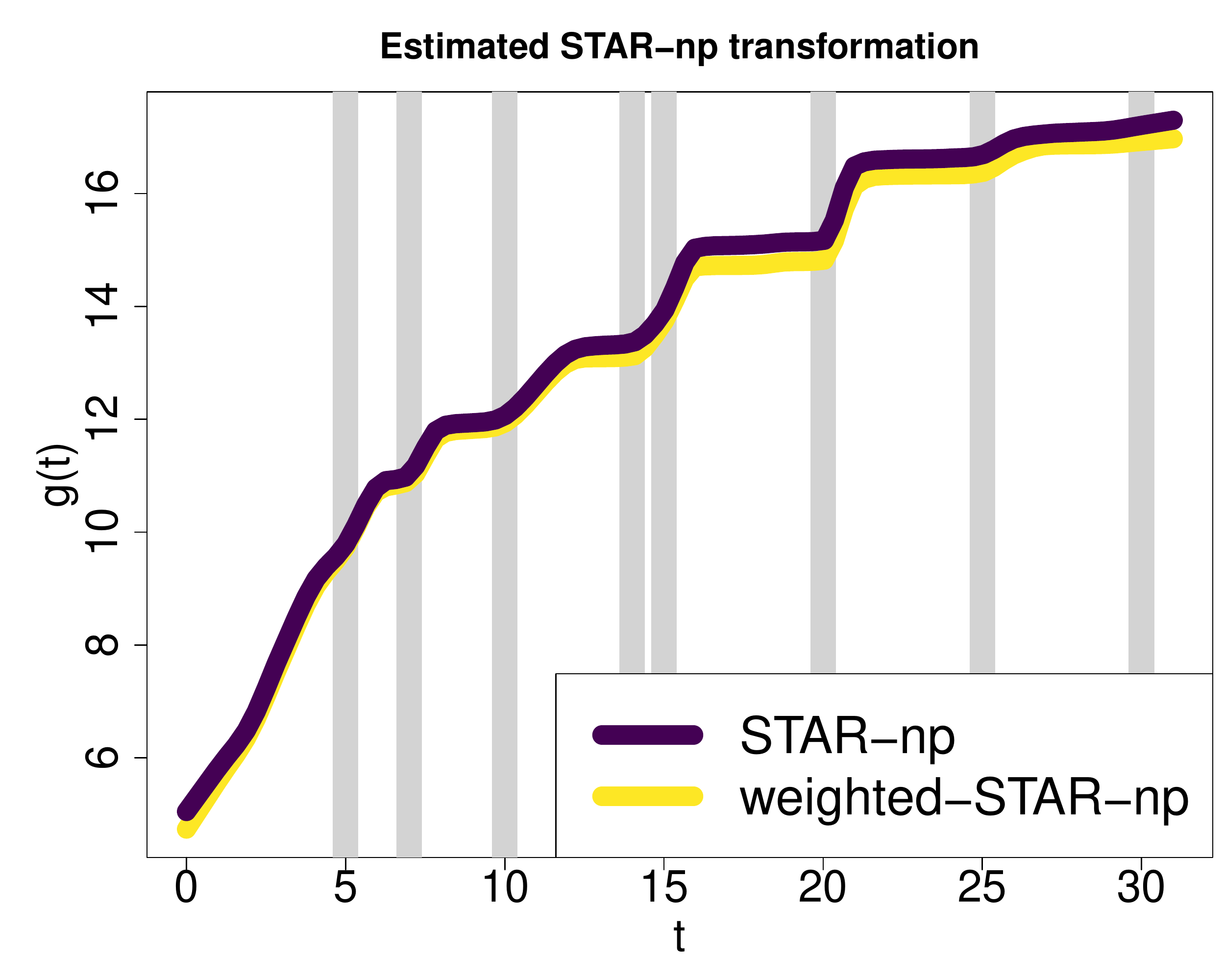}
\end{center}
\caption{\small Nonparametric estimate of the transformation $\widehat g$ along with the survey-weighted version. The heaping points are annotated. 
The most prominent changes in slope occur around $t \in \{7, 10, 15, 20\}$ and correspond to the primary heaping points observed in the PMF of $y$ (Figure~\ref{fig:pred}). The unweighted and weighted estimates are similar. Note that for general use, the unweighted version is preferred.
\label{fig:star-g}}
\end{figure}

The estimated nonparametric transformation \eqref{interpo} is presented in Figure~\ref{fig:star-g} along with the survey-weighted alternative. The most prominent changes in slope occur around $t \in \{7, 10, 15, 20\}$, which notably correspond to the primary heaping points observed in the PMF of $y$ (Figure~\ref{fig:pred}). This effect is by design: the nonparametric transformation \eqref{interpo} converts the empirical CDF of $y$ into a suitable transformation for the \STAR regression model  \eqref{spn}--\eqref{round}. Informally,  the nonparametric transformation ``stretches out" these heaps in the latent $z^*$-space for the continuous regression model \eqref{spn}. There is substantial agreement between the weighted and unweighted estimators.

For completeness, Table~\ref{tab:pvals} compares the marginal $p$-values among Gauss-log, ZIP, NegBin, ZINB, and the \STAR models. We note that this is not our recommended procedure for variable selection (see the AIC and BIC selections in Figure~\ref{fig:coef}) and that these $p$-values have not been corrected for survey sampling or multiple comparisons. However, these statistics provide useful summaries of the degree of inferential agreement among the competing models.   In general, \STARp-np and \STARp-sqrt agree with each other and with Gauss-log, although Gauss-log is not a valid count regression model. The alternative count data models---which are  inadequate according to residual diagnostics---offer little to no consensus with each other or \STAR models. These incongruities persist for AIC and BIC backward elimination akin to Figure~\ref{fig:coef} (not shown). Furthermore, the alternative models fail to identify several key effects: for example, BIC backward elimination for ZINB does \emph{not} select age, diabetes, blood pressure, or recent nicotine use.



\begin{table}[h]
\centering
\begin{tabular}{rrrrrrr}
  \hline
 & Gauss-log & ZIP & NegBin & ZINB & \STARp-np & \STARp-sqrt \\ 
  \hline
Black & 0.005 & 0.129 & 0.089 & 0.680 & 0.008 & 0.008 \\ 
  Hispanic & 0.006 & 0.129 & 0.088 & 0.719 & 0.010 & 0.011 \\ 
  RaceOther & 0.375 & 0.024 & 0.635 & 0.596 & 0.424 & 0.429 \\ 
  Female & 0.000 & 0.000 & 0.000 & 0.048 & 0.000 & 0.000 \\ 
  Age & 0.000 & 0.511 & 0.047 & 0.893 & 0.000 & 0.000 \\ 
  Widow/Divorce/Sep & 0.261 & 0.000 & 0.218 & 0.101 & 0.233 & 0.213 \\ 
  MaritalSingle & 0.487 & 0.000 & 0.261 & 0.127 & 0.420 & 0.381 \\ 
  High Income & 0.054 & 0.008 & 0.732 & 0.232 & 0.032 & 0.042 \\ 
  Low Income & 0.000 & 0.000 & 0.000 & 0.003 & 0.000 & 0.000 \\ 
  Education= HS & 0.820 & 0.043 & 0.813 & 0.413 & 0.995 & 0.953 \\ 
  Education$>$ HS & 0.544 & 0.011 & 0.408 & 0.412 & 0.587 & 0.562 \\ 
  HeavyDrinker & 0.000 & 0.000 & 0.006 & 0.035 & 0.000 & 0.000 \\ 
  HighBP & 0.000 & 0.000 & 0.007 & 0.110 & 0.000 & 0.000 \\ 
  HighChol & 0.089 & 0.976 & 0.160 & 0.950 & 0.084 & 0.085 \\ 
  Diabetes & 0.007 & 0.000 & 0.021 & 0.053 & 0.006 & 0.005 \\ 
  BMI & 0.493 & 0.005 & 0.555 & 0.311 & 0.533 & 0.519 \\ 
  Cholesterol & 0.872 & 0.497 & 0.673 & 0.730 & 0.768 & 0.790 \\ 
  UseNicotine & 0.001 & 0.000 & 0.025 & 0.169 & 0.004 & 0.004 \\ 
  UsedMarijuana & 0.019 & 0.072 & 0.159 & 0.703 & 0.012 & 0.015 \\ 
  UsedHardDrug & 0.059 & 0.002 & 0.271 & 0.384 & 0.055 & 0.048 \\ 
  Uninsured & 0.043 & 0.713 & 0.285 & 0.756 & 0.038 & 0.045 \\ 
  svypsu2 & 0.222 & 0.397 & 0.325 & 0.554 & 0.246 & 0.240 \\ 
  svypsu3 & 0.514 & 0.000 & 0.557 & 0.102 & 0.720 & 0.734 \\ 
   \hline
\end{tabular}
\caption{\small Marginal $p$-values for each regression coefficient ($H_{0j}: \beta_j = 0$ for  $j=1,\ldots,p$) under various models. There is strong agreement among \STARp-np, \STARp-sqrt, and Gauss-log.
\label{tab:pvals}}
\end{table}

\section{Simulation study}\label{sims}
The \STAR models are evaluated for point prediction,  distributional fitness, and hypothesis testing using simulated data. Among \STAR linear models, we compare the smoothed empirical CDF from Section~\ref{learn-g} (\STARp-np), Box-Cox with estimated $\lambda$ (\STARp-bc), and Box-Cox with fixed $\lambda = 1/2$ (\STARp-sqrt). Competing methods are Negative Binomial regression (NegBin), Poisson regression (Poisson), and Gaussian regression on $\log(y+1)$  (Gauss-log). 

We consider two count-valued data-generating processes: (i) \STAR data (Mixture-CDF) designed to match the empirical features of \texttt{DaysMentHlthNotGood} and (ii) Negative Binomial data with overdispersion.  For Mixture-CDF data, we specify the rounding operator to support $y \in \{0,\ldots, 30\}$. The transformation is constructed by smoothing the CDF of a random variable  that places $1/2$ mass on a $\mbox{Poisson}(10)$ distribution, $1/4$ mass uniformly on the set of heaps $\{5, 10, \ldots, 25\}$, and $1/4$ mass uniformly on the set of boundaries $\{0, 30\}$. The latent data-generating model \eqref{spn} is completed by fixing $\sigma = 0.7$. The Negative Binomial data are generated independently from  $ y_i \sim \mbox{NB}\{r^*,  \lambda_i^*/(r^* + \lambda_i^*) \}$ with expectation $\lambda_i^*$ and variance $\lambda_i^*(1 + \lambda_i^*/r^*)$. We specify $\log( \lambda_i^* )= \bm x_i'\bm \beta^*$ and set  $r^* = 3$ to include overdispersion. 

In both cases, we use $n = 500$ (see the supplement for $n=100$) 
and $p = 10$ with covariates generated from marginal standard normal distributions with $\mbox{Cor}(x_{i,j}, x_{i,j'}) = (0.75)^{|j - j'|}$. The $p$ columns are randomly permuted and augmented with an intercept. The true regression coefficients are defined by the intercept $\beta_0^* = \log(1.5)$, signals $\beta_j^* = \log(1.25)$ for  $j=1,\ldots, p/2 = 5$, and the remaining coefficients fixed at zero. The simulations are repeated for 100 iterations.

\begin{figure}[h!]
\begin{center}
\includegraphics[width=.49\textwidth]{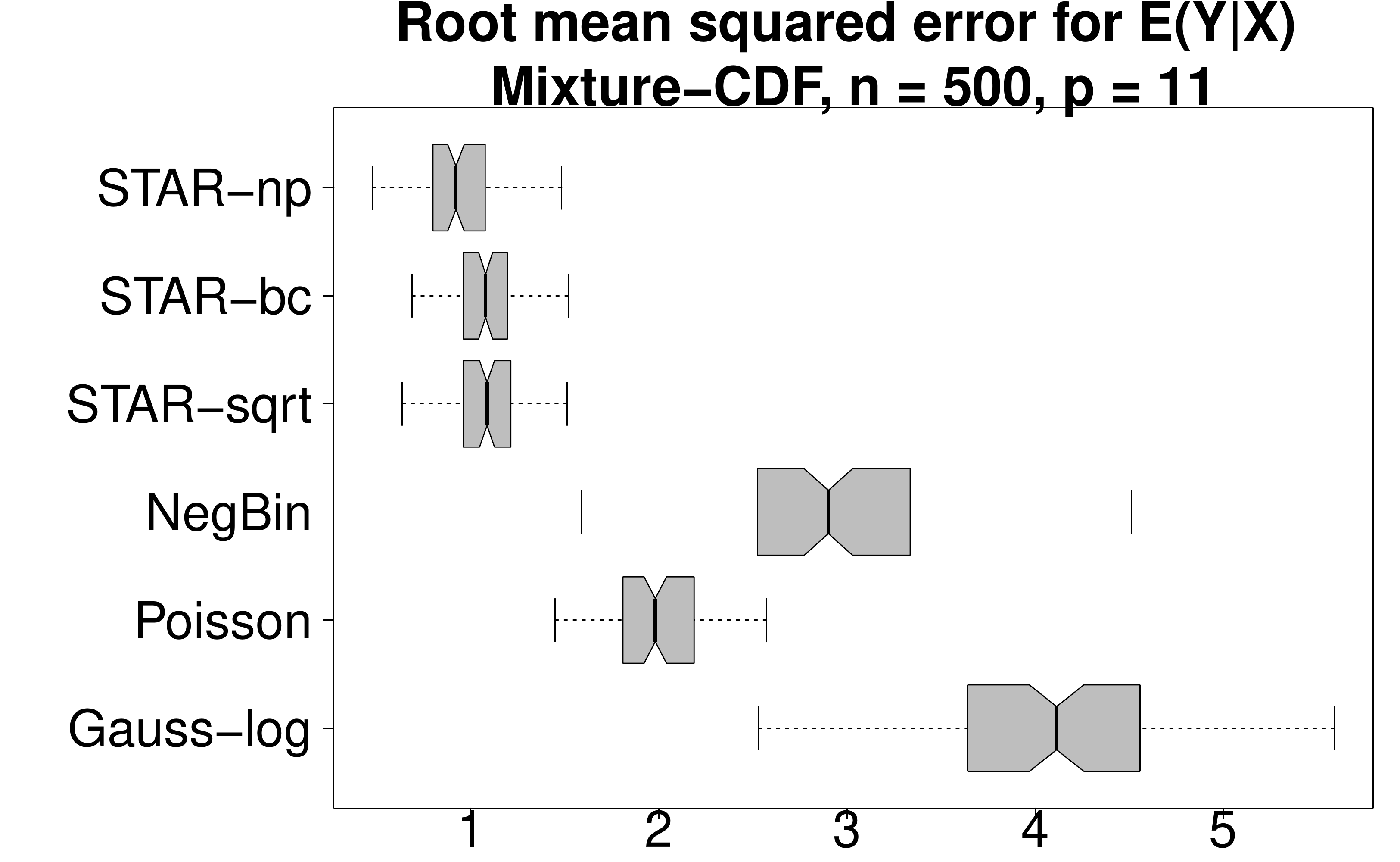}
\includegraphics[width=.49\textwidth]{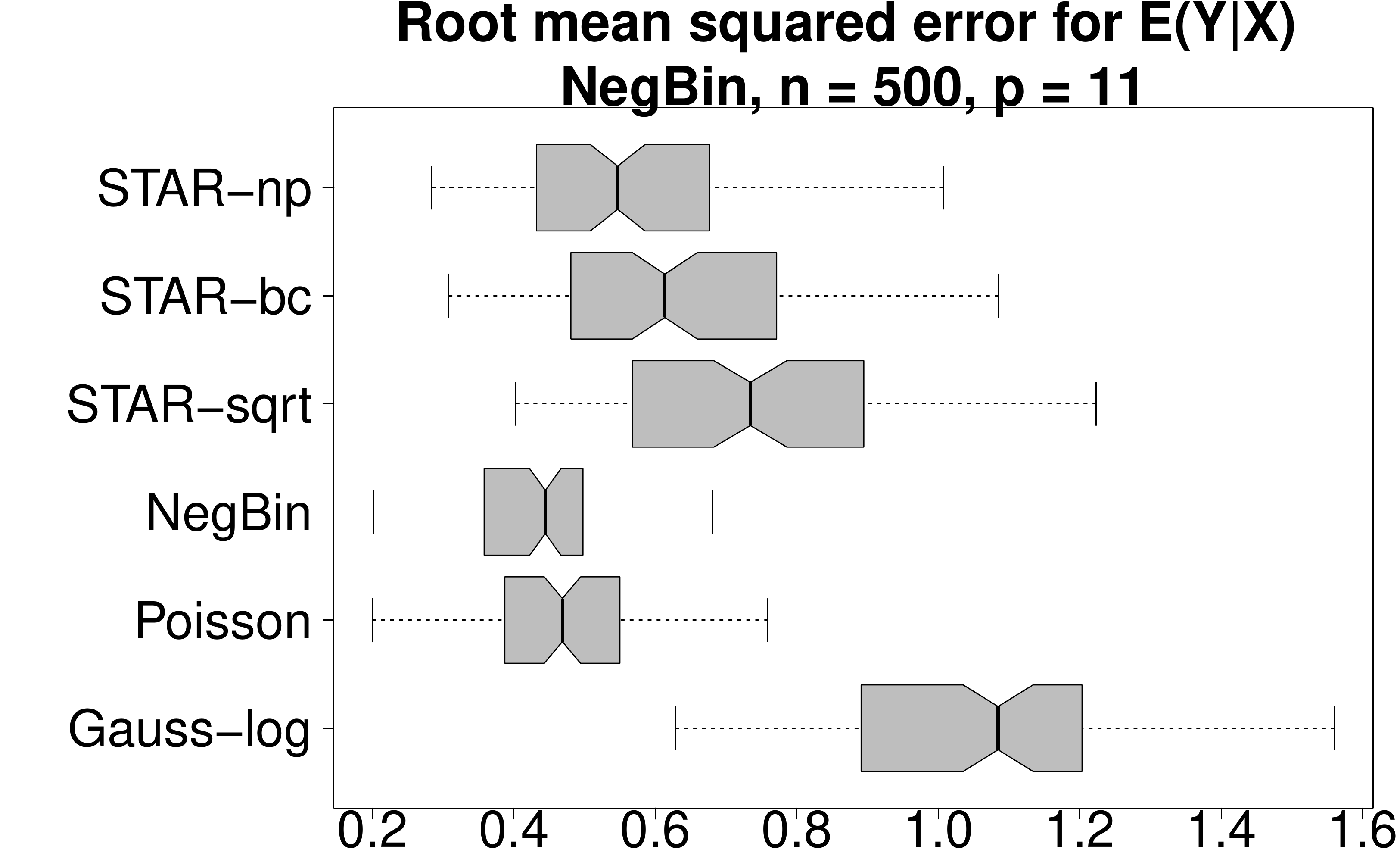}

\includegraphics[width=.49\textwidth]{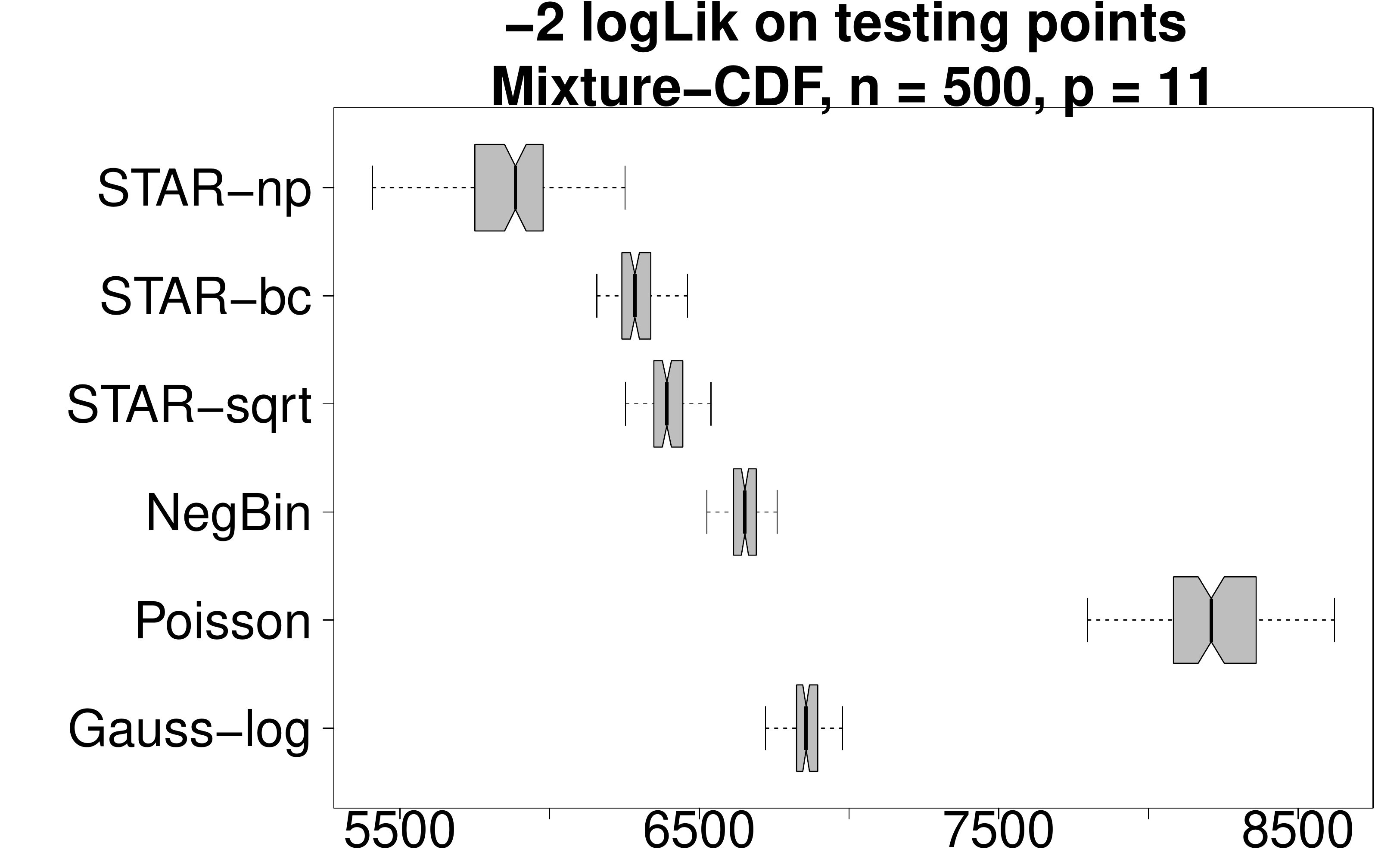}
\includegraphics[width=.49\textwidth]{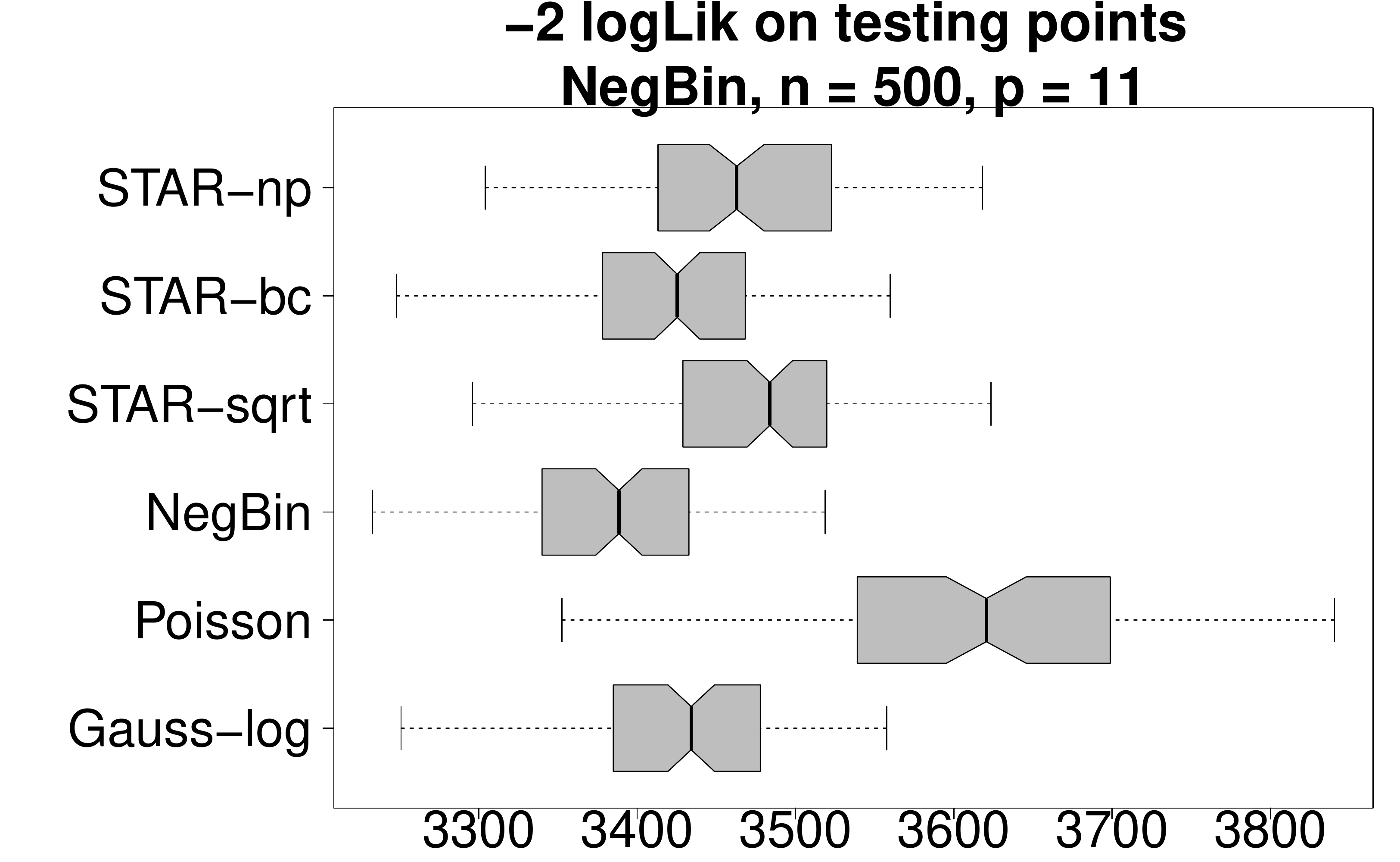}
\end{center}
\caption{\small Simulation results for Mixture-CDF data (left) and Negative Binomial data (right) evaluated for point prediction (RMSE,  top) and distributional fitness (-2 log-likelihood, bottom). \STAR and NegBin are best for their respective data-generating processes, but \STAR is surprisingly competitive for both point predictions and distributional accuracy with Negative Binomial data. 
\label{fig:sims-500}}
\end{figure}

Point prediction and distributional fitness are summarized in Figure~\ref{fig:sims-500}. Point prediction is evaluated using root mean squared error (RMSE) on $\mathbb{E}(y_i | \bm x_i)$, while distributional fitness is evaluated by computing -2 log-likelihood on a test set of 1000 points. The log-likelihood evaluations use each model likelihood (e.g., Poisson) with the parameters fixed at the MLEs. As expected, the \STAR models demonstrate vastly superior point prediction and distributional accuracy for the data exhibiting zero-inflation, boundedness, and heaping (Mixture-CDF). Similarly, the Negative Binomial data is best modeled by NegBin; however, \STAR is surprisingly competitive for both point predictions and distributional accuracy. In both cases, the Poisson model is among the best for point prediction yet the clear worst for distributional fitness. Note that Gauss-log is included in the likelihood comparisons but is not a coherent model for count data.

Table~\ref{table:tests-500} reports the Type I error and power at $\alpha = 0.10$ for testing the null hypothesis $H_{0j}: \beta_j = 0$ against the two-sided alternative. These values are averaged across $j=1,\ldots, p$ and all simulations. Most notably, the Poisson model is unable to control Type I error at the designated level. Despite the distinct data-generating processes, \STAR and NegBin perform similarly along with Gauss-log, with each method slightly favored under its own data-generating process.

\begin{table}[h] \small
\centering
\begin{tabular}{rrrrrrrr}
  \hline
 & & Gauss-log & Poisson & NegBin & \STARp-sqrt & \STARp-bc & \STARp-np \\ 
  \hline
Mixture-CDF
& Type I Error & 0.086 & 0.412 & 0.086 & 0.102 & 0.114 & 0.110 \\ 
&   Power & 0.896 & 0.974 & 0.916 & 0.938 & 0.938 & 0.948 \\

NegBin
&        Type I Error & 0.078 & 0.252 & 0.086 & 0.084 & 0.082 & 0.080 \\ 
 & Power & 0.830 & 0.940 & 0.882 & 0.854 & 0.844 & 0.838 \\
        
     \hline
\end{tabular}
\caption{\small Type I error rates and power for hypothesis testing in the linear model at the $\alpha = 0.10$ level for Mixture-CDF and Negative Binomial data. \STAR and NegBin perform similarly, while Poisson cannot control Type I error at the designated level. \label{table:tests-500}}
\end{table}

\section{Discussion}\label{discuss} 
Motivated by health questionnaire data,  we introduced semiparametric estimation and inference for count data regression. Our simultaneous transformation and rounding (\STARp) framework  ``sanitizes" continuous data models for use with discrete data exhibiting a variety of distributional phenomena.  Relative to existing count regression models, \STAR is most advantageous in the presence of challenging distributional features, especially heaping and boundedness (or censoring), as well as zero-inflation and over/underdispersion. Notably, \STAR regression is capable of modeling these features without adding more layers or parameters to the model. Instead, the modeling flexibility is enabled by the transformation and rounding operators: the transformation is estimated nonparametrically while the rounding operator ensures the correct (discrete)  support for the data. By design, \STAR leverages existing continuous data models and estimation algorithms: within our EM algorithm, a (weighted) least squares solution is sufficient to produce \STAR maximum likelihood estimators. Likelihood-based inference provides hypothesis testing, confidence intervals, variable selection via information criteria, and regression diagnostics. Simulation results confirmed the utility of the regression modeling approach under multiple designs. 

\STAR regression models were applied to self-reported mental health. While the mental health questionnaire responses presented challenges for existing count regression models, \STAR was capable of providing an adequate model fit that demonstrated substantial improvements in both likelihood-based comparisons and visual diagnostics. Using a \STAR linear model with a nonparametric transformation, we estimated effects and confidence intervals, incorporated adjustments for survey weights, selected variables (via AIC and BIC backward-elimination), and tested hypotheses for a broad set of demographic, socioeconomic, behavioral, and health-related covariates.

We emphasize that our results are neither causal nor direct measurements of poor mental health, but rather estimated associations with \emph{self-reported} mental health. In particular, our estimated effects (Figure~\ref{fig:coef}) do not directly imply that certain groups (e.g., Black or Hispanic individuals) are less likely to  \emph{experience} poor mental health  (all else equal), but rather that they are less likely to \emph{report} poor mental health (all else equal). Nonetheless, self-reporting of poor mental health remains an independently interesting topic; future research to decouple experienced and self-reported poor mental health would be valuable and complementary.

Lastly, we have used survey-weighted \STAR (Section~\ref{apps}) to visualize the impact of the NHANES sampling design on point estimation. However, valid uncertainty quantification and inference for general survey data remain open challenges for \STAR and other regression models \citep{Lumley2017}.

\noindent{\bf Supporting Information:} proofs of key results, details on the estimation algorithm, additional simulation results, and \texttt{R} code to reproduce the NHANES data pre-processing, NHANES data modeling and analysis, and simulated data comparisons are  online. An \texttt{R} package for \STAR regression is available: \url{https://github.com/drkowal/rSTAR}.



\bibliographystyle{apalike}
\bibliography{refs}

\begin{thebibliography}{}

\bibitem[Canale and Dunson, 2011]{canale2011bayesian}
Canale, A. and Dunson, D.~B. (2011).
\newblock {Bayesian kernel mixtures for counts}.
\newblock {\em Journal of the American Statistical Association},
  106(496):1528--1539.

\bibitem[Canale and Dunson, 2013]{canale2013nonparametric}
Canale, A. and Dunson, D.~B. (2013).
\newblock {Nonparametric Bayes modelling of count processes}.
\newblock {\em Biometrika}, 100(4):801--816.

\bibitem[Casella and Berger, 2002]{casella2002statistical}
Casella, G. and Berger, R.~L. (2002).
\newblock {\em {Statistical Inference}}, volume~2.
\newblock Duxbury Pacific Grove, CA.

\bibitem[Dempster et~al., 1977]{dempster1977maximum}
Dempster, A.~P., Laird, N.~M., and Rubin, D.~B. (1977).
\newblock {Maximum Likelihood from Incomplete Data Via the EM Algorithm}.
\newblock {\em Journal of the Royal Statistical Society: Series B
  (Methodological)}, 39(1):1--22.

\bibitem[Dunn and Smyth, 1996]{dunn1996randomized}
Dunn, P.~K. and Smyth, G.~K. (1996).
\newblock {Randomized quantile residuals}.
\newblock {\em Journal of Computational and Graphical Statistics},
  5(3):236--244.

\bibitem[Dunson, 2005]{Dunson2005}
Dunson, D.~B. (2005).
\newblock {Bayesian semiparametric isotonic regression for count data}.
\newblock {\em Journal of the American Statistical Association},
  100(470):618--627.

\bibitem[Fritsch and Carlson, 1980]{Fritsch1980}
Fritsch, F.~N. and Carlson, R.~E. (1980).
\newblock {Monotone piecewise cubic interpolation}.
\newblock {\em SIAM Journal on Numerical Analysis}, 17(2):238--246.

\bibitem[Gelman, 2007]{Gelman2007}
Gelman, A. (2007).
\newblock {Struggles with survey weighting and regression modeling}.
\newblock {\em Statistical Science}, 22(2):153--164.

\bibitem[Herring et~al., 2004]{Herring2004}
Herring, A.~H., Dunson, D.~B., and Dole, N. (2004).
\newblock {Modeling the effects of a bidirectional latent predictor from
  multivariate questionnaire data}.
\newblock {\em Biometrics}, 60(4):926--935.

\bibitem[Ives, 2015]{ives2015testing}
Ives, A.~R. (2015).
\newblock {For testing the significance of regression coefficients, go ahead
  and log-transform count data}.
\newblock {\em Methods in Ecology and Evolution}, 6(7):828--835.

\bibitem[Kipnis et~al., 2009]{Kipnis2009}
Kipnis, V., Midthune, D., Buckman, D.~W., Dodd, K.~W., Guenther, P.~M.,
  Krebs‐Smith, S.~M., Subar, A.~F., Tooze, J.~A., Carroll, R.~J., and
  Freedman, L.~S. (2009).
\newblock {Modeling data with excess zeros and measurement error: application
  to evaluating relationships between episodically consumed foods and health
  outcomes}.
\newblock {\em Biometrics}, 65(4):1003--1010.

\bibitem[Klungs{\o}yr et~al., 2006]{Klungsoyr2006}
Klungs{\o}yr, O., Nyg{\aa}rd, J.~F., S{\o}rensen, T., and Sandanger, I. (2006).
\newblock {Cigarette smoking and incidence of first depressive episode: an
  11-year, population-based follow-up study}.
\newblock {\em American Journal of Epidemiology}, 163(5):421--432.

\bibitem[Kowal, 2021]{Kowal2020target}
Kowal, D.~R. (2021).
\newblock {Fast, Optimal, and Targeted Predictions using Parametrized Decision
  Analysis}.
\newblock {\em Journal of the American Statistical Association}.

\bibitem[Kowal and Canale, 2020]{Kowal2020a}
Kowal, D.~R. and Canale, A. (2020).
\newblock {Simultaneous Transformation and Rounding (STAR) Models for
  Integer-Valued Data}.
\newblock {\em Electronic Journal of Statistics}, 14(1):1744--1772.

\bibitem[Liu et~al., 2009]{Liu2009}
Liu, H., Lafferty, J., and Wasserman, L. (2009).
\newblock {The nonparanormal: Semiparametric estimation of high dimensional
  undirected graphs}.
\newblock {\em Journal of Machine Learning Research}, 10(10):2295--2328.

\bibitem[Lumley and Scott, 2017]{Lumley2017}
Lumley, T. and Scott, A. (2017).
\newblock {Fitting regression models to survey data}.
\newblock {\em Statistical Science}, pages 265--278.

\bibitem[McCullagh and Nelder, 1989]{mccullagh1989generalized}
McCullagh, P. and Nelder, J.~A. (1989).
\newblock {\em {Generalized Linear Models, Vol. 37 of Monographs on Statistics
  and Applied Probability}}.
\newblock Chapman and Hall, London.

\bibitem[Mirowsky and Ross, 1999]{Mirowsky1999}
Mirowsky, J. and Ross, C.~E. (1999).
\newblock {Well-being across the life course}.
\newblock {\em A handbook for the study of mental health}, pages 328--347.

\bibitem[Niu and Hoff, 2019]{Niu2019}
Niu, X. and Hoff, P.~D. (2019).
\newblock {Joint mean and covariance modeling of multiple health outcome
  measures}.
\newblock {\em The Annals of Applied Statistics}, 13(1):321.

\bibitem[O'Hara and Kotze, 2010]{o2010not}
O'Hara, R.~B. and Kotze, D.~J. (2010).
\newblock {Do not log-transform count data}.
\newblock {\em Methods in Ecology and Evolution}, 1(2):118--122.

\bibitem[Ortega and Corzine, 1990]{Ortega1990}
Ortega, S.~T. and Corzine, J. (1990).
\newblock {Socioeconomic status and mental disorders}.
\newblock {\em Research in community and mental health}, 6:149--182.

\bibitem[Pratt, 1981]{pratt1981concavity}
Pratt, J.~W. (1981).
\newblock {Concavity of the log likelihood}.
\newblock {\em Journal of the American Statistical Association},
  76(373):103--106.

\bibitem[Scheid and Wright, 2017]{Scheid2017}
Scheid, T.~L. and Wright, E.~R. (2017).
\newblock {\em {A handbook for the study of mental health: Social contexts,
  theories, and systems}}.
\newblock Cambridge University Press.

\bibitem[Seedat et~al., 2009]{Seedat2009}
Seedat, S., Scott, K.~M., Angermeyer, M.~C., Berglund, P., Bromet, E.~J.,
  Brugha, T.~S., Demyttenaere, K., Girolamo, G.~D., Haro, J.~M., and Jin, R.
  (2009).
\newblock {Cross-national associations between gender and mental disorders in
  the World Health Organization World Mental Health Surveys}.
\newblock {\em Archives of general psychiatry}, 66(7):785--795.

\bibitem[Sellers and Shmueli, 2010]{sellers2010flexible}
Sellers, K.~F. and Shmueli, G. (2010).
\newblock {A flexible regression model for count data}.
\newblock {\em The Annals of Applied Statistics}, pages 943--961.

\bibitem[Siegfried and Hothorn, 2020]{Siegfried2020}
Siegfried, S. and Hothorn, T. (2020).
\newblock {Count transformation models}.
\newblock {\em Methods in Ecology and Evolution}, 11(7):818--827.

\bibitem[Song et~al., 2017]{Song2017}
Song, X., Xia, Y., and Zhu, H. (2017).
\newblock {Hidden Markov latent variable models with multivariate longitudinal
  data}.
\newblock {\em Biometrics}, 73(1):313--323.

\bibitem[St-Pierre et~al., 2018]{st2018count}
St-Pierre, A.~P., Shikon, V., and Schneider, D.~C. (2018).
\newblock {Count data in biology--Data transformation or model reformation?}
\newblock {\em Ecology and evolution}, 8(6):3077--3085.

\bibitem[Tzourio et~al., 1999]{Tzourio1999}
Tzourio, C., Dufouil, C., Ducimeti{\`{e}}re, P., and Alp{\'{e}}rovitch, A.
  (1999).
\newblock {Cognitive decline in individuals with high blood pressure: a
  longitudinal study in the elderly}.
\newblock {\em Neurology}, 53(9):1948.

\bibitem[Valle et~al., 2019]{valle2019ordinal}
Valle, D., Toh, K.~B., Laporta, G.~Z., and Zhao, Q. (2019).
\newblock {Ordinal regression models for zero-inflated and/or over-dispersed
  count data}.
\newblock {\em Scientific reports}, 9(1):3046.

\bibitem[Warton et~al., 2016]{warton2016three}
Warton, D.~I., Lyons, M., Stoklosa, J., and Ives, A.~R. (2016).
\newblock {Three points to consider when choosing a LM or GLM test for count
  data}.
\newblock {\em Methods in Ecology and Evolution}, 7(8):882--890.

\bibitem[Wilks, 1938]{wilks1938large}
Wilks, S.~S. (1938).
\newblock {The large-sample distribution of the likelihood ratio for testing
  composite hypotheses}.
\newblock {\em The Annals of Mathematical Statistics}, 9(1):60--62.

\bibitem[Williams et~al., 2010a]{Williams2010race}
Williams, D.~R., Costa, M., and Leavell, J.~P. (2010a).
\newblock {Race and mental health: Patterns and challenges}.
\newblock {\em A handbook for the study of mental health: Social contexts,
  theories, and systems}, pages 268--290.

\bibitem[Williams et~al., 2010b]{Williams2010}
Williams, K., Frech, A., and Carlson, D.~L. (2010b).
\newblock {Marital status and mental health}.
\newblock {\em A handbook for the study of mental health: Social contexts,
  theories, and systems}, pages 306--320.

\end{thebibliography}

\begin{table}[h]
\centering
\begin{tabular}{m{6cm} m{7cm}} 
\textbf{Variable} &  \textbf{Values} \\
\hline
   \rowcolor{shade}
 \multicolumn{2}{l}{\textbf{Response variable:}} \\
\href{https://wwwn.cdc.gov/Nchs/Nhanes/2011-2012/HSQ_G.htm#HSQ480}{DaysMentHlthNotGood} &  $\{0, 1, \ldots, 30\}$ \\
 \hline
  \rowcolor{shade}
  \multicolumn{2}{l}{\textbf{Demographic and socioeconomic variables:}} \\
 \href{https://wwwn.cdc.gov/nchs/nhanes/2011-2012/demo_g.htm#RIAGENDR}{Gender} &  {\emph{Male} ($54\%$), Female ($46\%$)} \\
 \rowcolor{shade}
 \href{https://wwwn.cdc.gov/nchs/nhanes/2011-2012/demo_g.htm#RIDAGEYR}{Age (years)}  &   $\{20, \ldots, 59\}$  \\
 \href{https://wwwn.cdc.gov/nchs/nhanes/2011-2012/demo_g.htm#RIDRETH1}{Race$^*$} &  {\emph{White} ($40\%$), Black ($25\%$), Hispanic ($20\%$), Other ($15\%$)} \\
  \rowcolor{shade}
 \href{https://wwwn.cdc.gov/nchs/nhanes/2011-2012/demo_g.htm#DMDMARTL}{Marital Status$^*$} &{\text{\emph{Married/Partner} ($58\%$)}, \text{Widowed/Divorced/Separated ($14\%$)}, \text{Single ($28\%$)}}\\
 \href{https://wwwn.cdc.gov/nchs/nhanes/2011-2012/demo_g.htm#INDFMIN2}{Family Income Level$^*$} &  {Low (26\%), \emph{Middle (57\%)}, High (17\%)}\\
  \rowcolor{shade}
 \href{https://wwwn.cdc.gov/nchs/nhanes/2011-2012/demo_g.htm#DMDHREDU}{Education Level$^*$} & {\emph{$<$ HS} ($19\%$), = HS ($19\%$), $>$ HS ($62\%$)}\\
\href{https://wwwn.cdc.gov/Nchs/Nhanes/2011-2012/HIQ_G.htm#HIQ011}{Uninsured$^*$}&{Yes ($30\%$), \emph{No}  ($70\%$)} \\
 \hline
   \rowcolor{shade}
 \multicolumn{2}{l}{\textbf{Alcohol and drug use variables:}} \\
 \href{https://wwwn.cdc.gov/Nchs/Nhanes/2011-2012/ALQ_G.htm#ALQ151}{HeavyDrinker} & {Yes ($16\%$), \emph{No} ($84\%$)}\\
   \rowcolor{shade}
 \href{https://wwwn.cdc.gov/Nchs/Nhanes/2011-2012/SMQRTU_G.htm#SMQ680}{UseNicotine} & {Yes ($31\%$), \emph{No}  ($69\%$)}\\
 \href{https://wwwn.cdc.gov/Nchs/Nhanes/2011-2012/DUQ_G.htm#DUQ200}{UsedMarijuana} & {Yes ($60\%$), \emph{No}  ($40\%$)}\\
   \rowcolor{shade}
 \href{https://wwwn.cdc.gov/Nchs/Nhanes/2011-2012/DUQ_G.htm#DUQ240}{UsedHardDrug} & {Yes ($20\%$), \emph{No}  ($80\%$)}\\
\hline
\multicolumn{2}{l}{\textbf{Health-related variables:}} \\
  \rowcolor{shade}
\href{https://wwwn.cdc.gov/nchs/nhanes/2011-2012/BMX_G.htm#BMXBMI}{Body Mass Index (BMI, kg/$m^2$)} & $[13.6, 69.0]$  \\
\href{https://wwwn.cdc.gov/nchs/nhanes/2011-2012/TCHOL_G.htm#LBXTC}{Cholesterol (total, mg/dL)} & $[59.0, 69.0]$ \\
  \rowcolor{shade}
\href{http://data.nber.org/nhanes/2011-2012/BPQ_G.htm}{HasHighBP (BPQ020 at link) }& {Yes ($25\%$), \emph{No}  ($75\%$)} \\
\href{http://data.nber.org/nhanes/2011-2012/BPQ_G.htm}{HasHighChol (BPQ080 at link) }& {Yes ($25\%$), \emph{No}  ($75\%$)} \\
  \rowcolor{shade}
\href{https://wwwn.cdc.gov/Nchs/Nhanes/2011-2012/DIQ_G.htm#DIQ010}{HasDiabetes$^*$}&{Yes ($9\%$), \emph{No}  ($91\%$)} \\
 \hline
\multicolumn{2}{l}{\textbf{Survey design variables:}} \\
\rowcolor{shade}
\href{https://wwwn.cdc.gov/Nchs/Nhanes/2011-2012/DEMO_G.htm#WTMEC2YR}{Sampling Weights}& $[0, 222579.8]$ \\ [1ex] 
\href{https://wwwn.cdc.gov/Nchs/Nhanes/2011-2012/DEMO_G.htm#SDMVPSU}{Primary Sampling Units}&  \emph{1} (48\%), 2 (44\%), 3 (8\%)
\\ [1ex]
\end{tabular}
\caption{\small Variables in the analysis dataset with hyperlinks to the online NHANES descriptions. The baseline categories for one-hot/dummy encoding are italicized. The continuous variables (Age, BMI, and Cholesterol) are centered and scaled prior to model fitting.  Annotated variables ($*$) include minor modifications (e.g., collapsed categories) from the original NHANES variables; see the online Data Dictionary for details. \label{tab:data}}
\end{table}

\end{document}